\documentclass[10pt,twocolumn,letterpaper]{article}

\usepackage[pagenumbers]{cvpr} 

\usepackage[dvipsnames]{xcolor}

\usepackage{times}
\usepackage{epsfig}
\usepackage{graphicx}
\usepackage{amsmath}
\usepackage{amssymb}

\usepackage{tikz}
\usepackage{graphicx}
\usepackage{amsmath}
\usepackage{amssymb}
\usepackage{booktabs}
\usepackage{textcomp}
\usepackage{comment}
\usepackage{adjustbox}
\usepackage{float}

\usepackage{svg}
\usepackage{tabularx}
\usepackage{pifont}
\newcommand{\cmark}{\ding{51}}%
\usepackage{multirow}
\usepackage{float}
\usepackage{subcaption}

\usepackage[utf8]{inputenc} 
\usepackage[T1]{fontenc}    
\usepackage{url}            
\usepackage{booktabs}       
\usepackage{amsfonts}       
\usepackage{nicefrac}       
\usepackage{microtype}      
\usepackage{xcolor}         

\definecolor{cvprblue}{rgb}{0.21,0.49,0.74}
\usepackage[pagebackref,breaklinks,colorlinks,citecolor=cvprblue]{hyperref}

\def\paperID{246} 
\def\confName{3DV\xspace}
\def\confYear{2026\xspace}

\title{PromptVFX: Text-Driven Fields\\for Open-World 3D Gaussian Animation}

\author{%
  Mert Kiray$^{*,1,2,3}$ \quad
  Paul Uhlenbruck$^{*,1}$ \quad
  Nassir Navab$^{1,2}$ \quad
  Benjamin Busam$^{1,2,3}$ \\[1ex]
  \begin{tabular}{ccc}
    $^{1}$Technical University of Munich &
    $^{2}$Munich Center for Machine Learning (MCML) &
    $^{3}$Obsphera
  \end{tabular}
}

\begin{document}
\makeatletter
\g@addto@macro\@maketitle{
  \vspace{-30pt}
  \begin{figure}[H]
  \setlength{\linewidth}{\textwidth}
  \setlength{\hsize}{\textwidth}
  \centering
    \includegraphics[width=\linewidth]{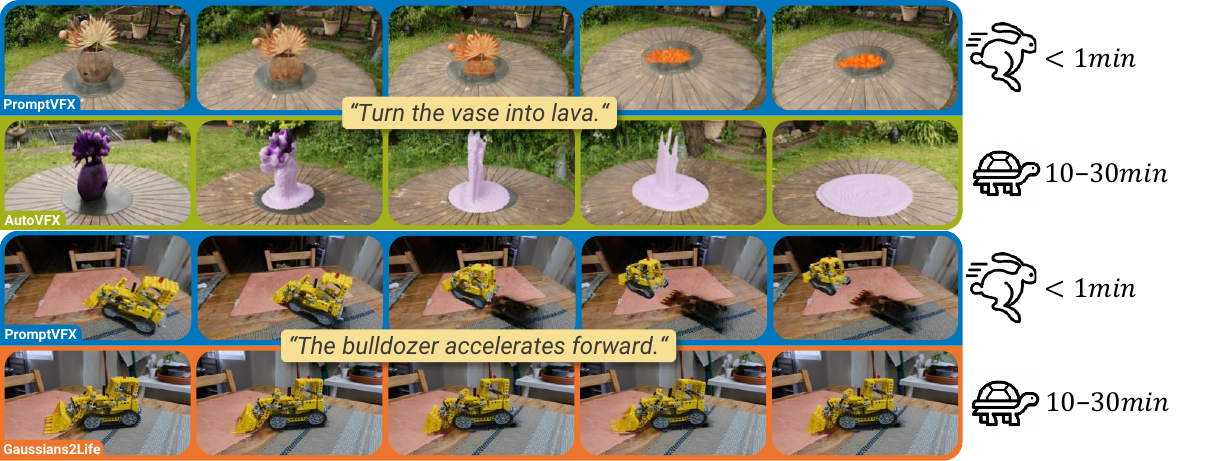}
 \caption{{\bf Comparison of PromptVFX with existing text-driven 3D animation methods.} Our approach generates high-quality animations in seconds, whereas others require 30× more time due to diffusion or physics-based simulation. Exact prompts are provided in the supplementary material.
    }
    \label{fig:teaser}
	\end{figure}
}
\makeatother

\maketitle

\begingroup
\renewcommand{\thefootnote}{\fnsymbol{footnote}}
\setcounter{footnote}{0}
\footnotetext[1]{Authors contributed equally to this work.}
\endgroup

\begin{abstract}
Visual effects (VFX) are key to immersion in modern films, games, and AR/VR. Creating 3D effects requires specialized expertise and training in 3D animation software and can be time consuming. Generative solutions typically rely on computationally intense methods such as diffusion models which can be slow at 4D inference. We reformulate 3D animation as a field prediction task and introduce a text-driven framework that infers a time-varying 4D flow field acting on 3D Gaussians. By leveraging large language models (LLMs) and vision-language models (VLMs) for function generation, our approach interprets arbitrary prompts (e.g., “make the vase glow orange, then explode”) and instantly updates color, opacity, and positions of 3D Gaussians in real time. This design avoids overheads such as mesh extraction, manual or physics-based simulations and allows both novice and expert users to animate volumetric scenes with minimal effort on a consumer device even in a web browser~\footnote{Web Demo: \url{http://promptvfx.duckdns.org}}. Experimental results show that simple textual instructions suffice to generate compelling time-varying VFX, reducing the manual effort typically required for rigging or advanced modeling. We thus present a fast and accessible pathway to language-driven 3D content creation that can pave the way to democratize VFX further. Code available at \url{https://obsphera.github.io/promptvfx/}.
\end{abstract}
\vspace{-1em}    
\begin{table*}[t]
\caption{{\bf Comparison of visual editing methods.} The table shows existing methods sorted by features regarding input/output, interfaces, effect capacities, and system properties. PromptVFX (Ours) is the only method that simultaneously fullfills all features while allowing for interactive user refinement.}
    \vspace{-10pt}
    \centering
    \resizebox{\linewidth}{!}{%
\begin{tabular}{l|cclccc|cccc|cccc}
\toprule
\multicolumn{1}{l}{} & \multicolumn{6}{c|}{Input \& Output} & \multicolumn{4}{c|}{Editing Capacities} & \multicolumn{4}{c}{Properties} \\\midrule
    Method & {\begin{tabular}[c]{@{}c@{}}Real World\\Video Editing\end{tabular}} & \begin{tabular}[c]{@{}c@{}}Free-Viewpoint\\Rendering\end{tabular} & {\begin{tabular}[c]{@{}c@{}}Editing\\Interface\end{tabular}} & \begin{tabular}[c]{@{}c@{}}Viewpoint\\Consistent\end{tabular} & \begin{tabular}[c]{@{}c@{}}Continuous\\Time Rep.\end{tabular} & \begin{tabular}[c]{@{}c@{}}Open-world \\ Query\end{tabular} & {\begin{tabular}[c]{@{}c@{}}Interactive\\Refinement\end{tabular}} & {\begin{tabular}[c]{@{}c@{}}Appearance\\Change\end{tabular}} & {\begin{tabular}[c]{@{}c@{}}Object\\Animations\end{tabular}} & \begin{tabular}[c]{@{}c@{}}Particle\\Effects\end{tabular} & {\begin{tabular}[c]{@{}c@{}}Zero-\\Shot\end{tabular}} & {\begin{tabular}[c]{@{}c@{}}No Complex\\Preprocessing\end{tabular}} & {\begin{tabular}[c]{@{}c@{}}Diffusion\\Model Free\end{tabular}} & {\begin{tabular}[c]{@{}c@{}}Animation\\Software Free\end{tabular}} \\\midrule

Visual Programming~\cite{gupta2023visual} & {\cmark} &  & {Language} &  &  & \cmark &  & {\cmark} & {} & {} & {\cmark} &  &  & \textbf{} \\
FRESCO~\cite{yang2024fresco} & {\cmark} &  & {Language} &  &  & \cmark &  & {\cmark} & {} & {} & {\cmark} & {\cmark} &  & {\cmark} \\
ClimateNeRF~\cite{Li2023ClimateNeRF} & {\cmark} & \cmark & {Scripts} & {\cmark} &  &  &  & {\cmark} & {} & \cmark &  & {\cmark} &  &  \\
GaussianEditor~\cite{chen2024gaussianeditor} & {\cmark} & \cmark & {GUI} & {\cmark} &  & \cmark &  & {\cmark} & {} &  &  & {\cmark} &  & {\cmark} \\
Gaussian Grouping~\cite{ye2024gaussian} & {\cmark} & \cmark & {GUI} & {\cmark} &  & \cmark &  & {\cmark} & {} &  &  &  &  & {\cmark} \\
PhysGaussian~\cite{xie2024physgaussian} & {\cmark} & \cmark & {GUI} & {\cmark} & {\cmark} &  & {\cmark} & {} & {} &  &  &  & {\cmark} &  \\
VR-GS~\cite{jiang2024vr} & {\cmark} & \cmark & {GUI} & {\cmark} & {\cmark} &  & {\cmark} & {} & {} &  &  &  &  &  \\
Gaussian Splashing~\cite{feng2024splashing} & {\cmark} & \cmark & {GUI} & {\cmark} &  &  &  & {} & {} & \cmark &  &  & {\cmark} &  \\
DMRF~\cite{qiao2023dynamic} & {\cmark} & \cmark & {GUI} & {\cmark} &  &  &  & {} & {\cmark} & \cmark &  & {\cmark} & {\cmark} & {\cmark} \\
Instruct-N2N~\cite{instructnerf2023} & {\cmark} & \cmark & {Language} & {\cmark} &  & \cmark &  & {\cmark} & {} &  &  &  &  & {\cmark} \\
DGE~\cite{chen2024dge} & {\cmark} & \cmark & {Language} & {\cmark} &  & \cmark &  & {\cmark} & {} &  &  &  &  & {\cmark} \\
Chat-Edit-3D~\cite{fang2024chat} & {\cmark} & \cmark & {Language} & {\cmark} &  & \cmark & {\cmark} & {\cmark} & {} &  &  & {\cmark} &  & {\cmark} \\
Gaussians2Life~\cite{wimmer2025gaussianstolife} & {\cmark} & \cmark & {Language} & {\cmark} &  & \cmark &  & {\cmark} & {\cmark} &  &  & {\cmark} &  & {\cmark} \\
DreamGaussians4D~\cite{ren2023dreamgaussian4d} & {\cmark} & \cmark & {Language} & {\cmark} & {\cmark} & \cmark &  & {\cmark} & {\cmark} &  &  & {\cmark} &  & {\cmark} \\
AutoVFX~\cite{hsu2024autovfx} & {\cmark} & \cmark & {Language} & {\cmark} & {\cmark} & \cmark &  & {\cmark} & {\cmark} & \cmark &  &  & {\cmark} &  \\
PromptVFX (Ours) & \cmark & \cmark & {Language} & {\cmark} & {\cmark} & \cmark & {\cmark} & \cmark & \cmark & {\cmark} & {\cmark} & {\cmark} & {\cmark} & {\cmark} \\
\bottomrule
\end{tabular}%
}
\label{tab:related_works}
\end{table*}

\section{Introduction}
\label{sec:intro}

"Today, with computer-generated visual effects, everything is possible. So we've seen everything. If it can be imagined, it can be put on screen."\footnote{Gabriel Campisi. Producer, Screenwriter, Director, and Author.}
The creation of 3D animation is omnipresent, yet it requires specialized software and expert knowledge like the ones from visual effect studios Campisi typically works with for his Hollywood productions.

VFX artists rely on offline physical simulation software for precise motion and collision dynamics which are computationally expensive and prohibitively slow on consumer hardware~\cite{blender}. Recent trends in generative AI~\cite{rombach2022high,shi2022deep} make the creative process more accessible to a wider audience. The pipelines typically employ large diffusion models to generate or edit 3D scenes with text prompts~\cite{poole2022dreamfusion, wimmer2025gaussianstolife}.
Computation cost and multi-view inconsistencies are obstacles on the way to realistic real-time interaction~\cite{shi2023zero123++,yang2024consistnet}.

To truly democratize 3D animation, a fast and intuitive system that permits open-ended text instructions with instant feedback is required. To this end, promising frameworks~\cite{hsu2024autovfx,instructnerf2023} have been created that integrate large (vision) language models (L(V)LMs) to parse user prompts. However, they equally depend on either physics-based simulations or diffusion training loops, preventing real-time editing. So far, there is no solution established that accepts any textual command (e.g., “rotate the statue, then change its color to gold in three seconds”) that immediately updates a 3D scene without offline simulation or optimization.

In this paper, we propose \textbf{PromptVFX}, a text-driven framework for 3D animation that provides these features. Centered on Gaussian Splatting~\cite{kerbl20233d}, our system translates high-level textual instructions into time-varying transformations on a set of Gaussians without generating new geometry or launching physics simulators.
A time-varying field is applied to a Gaussian splatting scene ultimately changing centre positions, colour and opacity of individual Gaussians.
When instructed to “raise the vase by two meters,” the system directly modifies the Gaussian centre positions over time. This simple yet generic idea naturally supports real-time rendering such that the effect of a user prompt can be visualized within seconds rather than minutes or hours.

A key enabler of our approach is its open-text interface, powered by an LLM and optionally a LVLM for appearance-grounding. In contrast to methods that restrict users to a narrow command set or domain-specific scripts, we allow any natural language request (“make the tree shake, then fade to transparent”). The LLM parses these instructions into text-driven fields that change the position, color and opacitiy of scene parts. By pairing VFX-transformations with a robust foundation model that has been trained on a large body of data, we can capitalize on the fact that "we've seen everything" (Campisi) and create generic cinematic or stylized effects within a single pipeline without training new models or worrying about 3D consistency of hallucinated content.
Our contributions can be summarized as follows:

\begin{itemize}
  \item \textbf{Open world VFX with text driven fields.} We recast 3D animation as time varying transformations on Gaussian splats to enable volumetric effects without retraining or simulation.
  \item \textbf{Training-free 4D animations.} Our zero-shot workflow breaks natural language instructions into animation phases, generates transformation functions via LLM and refines them using visual language model feedback with no per scene training.
  \item \textbf{Real-time interactive editing.} Updating Gaussian parameters directly removes mesh extraction, diffusion loops or simulators. Animations are delivered in under one minute on a single GPU or in a browser allowing for instant user feedback.
\end{itemize}

\noindent
We demonstrate our method on various scenes and user prompts, showcasing dynamic, visually consistent animations generated within seconds of receiving text instructions in \autoref{fig:teaser}. Beyond accelerating the workflow for experienced artists, this real-time, open-text interface makes 3D animation accessible to a significantly broader audience. We strongly believe that our work can help unify intuitive language interfaces with efficient 3D animation.
\begin{figure*}[t!]
    \centering
    \includegraphics[width=\linewidth]{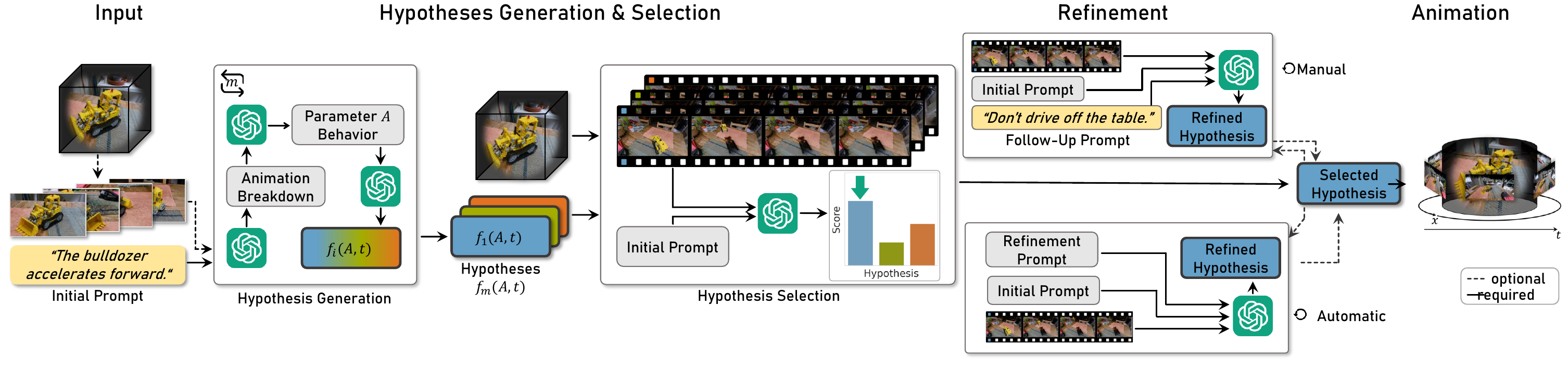}
    \vspace{-20pt}
    \caption{{\bf Overview of the PromptVFX pipeline.} Given a user-provided textual prompt, the system first decomposes it into structured animation phases. A large language model (LLM) then generates parametric functions that define the motion, color, and opacity changes of 3D Gaussians over time. To handle ambiguity, multiple animation hypotheses are generated and evaluated using a vision-language model (VLM) or user feedback. The selected animation is further refined through automatic and interactive text-based corrections, ensuring high-quality, real-time results. More details on prompt formulation and scoring are provided in the supplementary material. }
    \label{fig:pipeline}
\end{figure*}

\section{Related Work}
\label{sec:related_work}
PromptVFX builds upon recent advancements in generative models for text-driven content creation, 4D content generation, 3D editing, physics-based VFX, and interactive interfaces for scene manipulation. We review key contributions in these areas and position our work in relation to them. \autoref{tab:related_works} summarizes key features of related editing tools.

\noindent\textbf{Generative text-to-pixel models.}
Early generative models such as Denoising Diffusion Probabilistic Models (DDPM)~\cite{ho2020denoising} and latent diffusion models~\cite{rombach2022high} have demonstrated remarkable success in text-to-image generation. These approaches have been extended to video generation through text-to-video diffusion models~\cite{singer2023make,blattmann2023stable,blattmann2023align}, which generate temporally coherent image sequences but remain computationally expensive.
Diffusion models have also been adapted to synthesize images from different viewpoints of a scene~\cite{bahmani2024vd3d,he2024cameractrl}. However, enforcing multi-view consistency is not straightforward and generated results can change appearance under varied 3D viewpoint. Several works attempt to enforce view consistency through explicit constraints~\cite{xie2024sv4d,kuang2024collaborative,poole2022dreamfusion}.

\noindent\textbf{Dynamic 3D generation.}
The representation of 3D Gaussian splatting~\cite{kerbl20233d} has emerged as an efficient alternative to implicit neural representations like NeRF~\cite{mildenhall2021nerf}. Recent works extend Gaussian Splatting to dynamic 4D content~\cite{das2024neural,lin2024gaussianflow,Luiten2024dynamic,wu20244d}, enabling animated reconstructions with controlled deformations~\cite{jung2023deformable3dgaussiansplatting,Qian20243dgsavatar}. However, dynamic 3D content generation still lags behind video-based approaches in realism, typically focusing on isolated foreground or single objects~\cite{bahmani20244d,singer2023text,ling2024aligngaussians}.
Current approaches can require multiple hours of optimization for a few seconds of scene content~\cite{yu20254real}.

\noindent\textbf{Diffusion-based 3D editing.}
Editing static 3D scenes has been explored through diffusion-based refinement, such as InstructNeRF2NeRF~\cite{instructnerf2023}, which adapts the Pix2Pix paradigm~\cite{brooks2023instructpix2pix} for NeRF-based representations. Similar approaches exist for Gaussian splatting~\cite{igs2gs}, but these methods remain computationally expensive and require retraining for each edit with many 2D diffusion steps from various perspectives.

\noindent\textbf{Visual effects with simulations.}
Conventional VFX systems rely on physics engines for scene interactions, but their applications to generative 3D content remain limited. Existing approaches incorporate rigid-body physics~\cite{wei2024editable,xia2024video2game}, particle dynamics~\cite{feng2024splashing,Li2023ClimateNeRF}, and elastic deformations~\cite{zhang2024physdreamer}, all constraint to specific physical interactions. Spring-Gaus~\cite{zhong2024springgaus}, for instance, applies spring-mass dynamics to Gaussian splats in scenarios of objects falling onto a plane. More flexible approaches such as PhysGaussian~\cite{xie2024physgaussian} extend capabilities to a few simulation types.
An indirect way to facilitate object interactions is to first extract explicit meshes from the scene representation~\cite{yang2022neumesh,jiang2024vr,peng2022cageNeRF,xu2022deforming,yuang2022NeRFEditing}.
This allows to use existing mesh-based physics simulators at the cost of error propagation and compute.
Unlike these works, our method enables open-ended, user-driven animations without requiring physical constraints, explicit simulations or the preliminary extraction of object meshes.

\noindent\textbf{Interfaces: from code to language.}
Traditional 3D editing interfaces rely on GUI-based or script-driven tools~\cite{chen2024gaussianeditor,jiang2024vr,Li2023ClimateNeRF,qiu2024language,ye2024gaussian}, which often require expert knowledge. This prevents the models to be accessible to a wider user group. Recent approaches integrate LLMs for structured editing by splitting a complex task into smaller sub-elements~\cite{wang2024GaussianEditor,fang2024chat} at the cost of computational overhead. In contrast, our system offers an intuitive direct natural language interface allowing for democratization of 3D content editing while being fast enough to iterate during the creative process.

\noindent\textbf{Towards open-world 3D scene editing.}
Recent works have explored ways towards more flexible scene editing of Gaussian Splatting scenes. Animate3D~\cite{jiang2025animate3d} allows the animation of single assets but is constrained by dataset limitation and training. Gaussians2Life~\cite{wimmer2025gaussianstolife} extends DreamGaussian4D~\cite{ren2023dreamgaussian4d} by applying 2D diffusion models to generate motion sequences, which are then lifted to 3D, making them view-consistent at the cost of computational complexity. Our approach bypasses diffusion-based inference by directly modifying Gaussian parameters using field-based transformations which enables interactive updates.
Similarly, AutoVFX~\cite{hsu2024autovfx} integrates LLM-generated scripts into Blender-based simulations, producing highly realistic effects at the cost of compute. While producing highly realistic scenes for supported effects, this approach is limited to pre-defined VFX modules. In contrast, we apply transformations directly on Gaussian splats, providing a more flexible and efficient framework for text-driven animation.

While previous works have demonstrated powerful generative capabilities in 3D content creation, they often suffer from slow inference times, limited editability, or reliance on predefined physical models. Our method introduces a novel approach by reformulating 3D animation as a field prediction task, allowing interactive, open-world text-driven modifications without additional training or per-scene optimization.
\section{Methodology}
\label{sec:method}
Our goal is to enable real-time, text-driven animation of 3D scenes without relying on diffusion-based optimization or physics simulators. In this section, we describe our approach in four parts: 
\begin{enumerate}[label=(\arabic*)]
    \item representation of 3D scenes as Gaussian splats,
    \item formulation of continuous time-dependent animation fields,
    \item LLM-based translation of open-domain text into parametric fields, and
    \item pipeline details ensuring efficient, user-friendly performance.
\end{enumerate}
\autoref{fig:pipeline} provides an overview of our zero-shot pipeline.

\subsection{3D Representation via Gaussian Splatting}
\label{subsec:gaussian_repr}
We adopt a Gaussian splatting (3DGS) representation for each object or region in our scene. Concretely, an object is approximated by a set of elliptical Gaussians $\{G_i\}_{i=1}^n$, where each $G_i$ is defined by:
\begin{itemize}
    \item $\mu_i \in \mathbb{R}^3$: the center position in 3D,
    \item $\Sigma_i \in \mathbb{R}^{3 \times 3}$: a covariance (or scale/rotation) tensor,
    \item $\mathbf{c}_i \in \mathbb{R}^3$: an RGB color vector (or spherical harmonic coefficients for view dependence),
    \item $\alpha_i \in \mathbb{R}$: an opacity or density parameter.
\end{itemize}
This explicit point-based representation can be rendered in real time via fast rasterization~\cite{kerbl20233d}, and it enables direct manipulation of positions, colors, and opacities without requiring expensive mesh extraction. Thus, any user-specified animation can dynamically update the attributes $\text{\textbf{A}}$ with $\text{\textbf{A}}_i = \left(\mu_i, \mathbf{c}_i, \alpha_i\right)$ on the fly. 

To animate a specific object within the scene, we first identify its corresponding Gaussians. Our framework remains agnostic to the selection method, enabling seamless integration with automated 3D segmentation techniques. Methods such as LangSplat~\cite{qin2024langsplat}, Gaussian Grouping~\cite{gaussian_grouping}, or GARField~\cite{garfield2024} can be employed to localize objects within the scene and associate them with their respective Gaussians. In this work, we assume that the segmentation of the target object is provided, with its corresponding Gaussians pre-selected for animation.

\begin{figure*}[t]
    \centering
    \includegraphics[width=\linewidth]{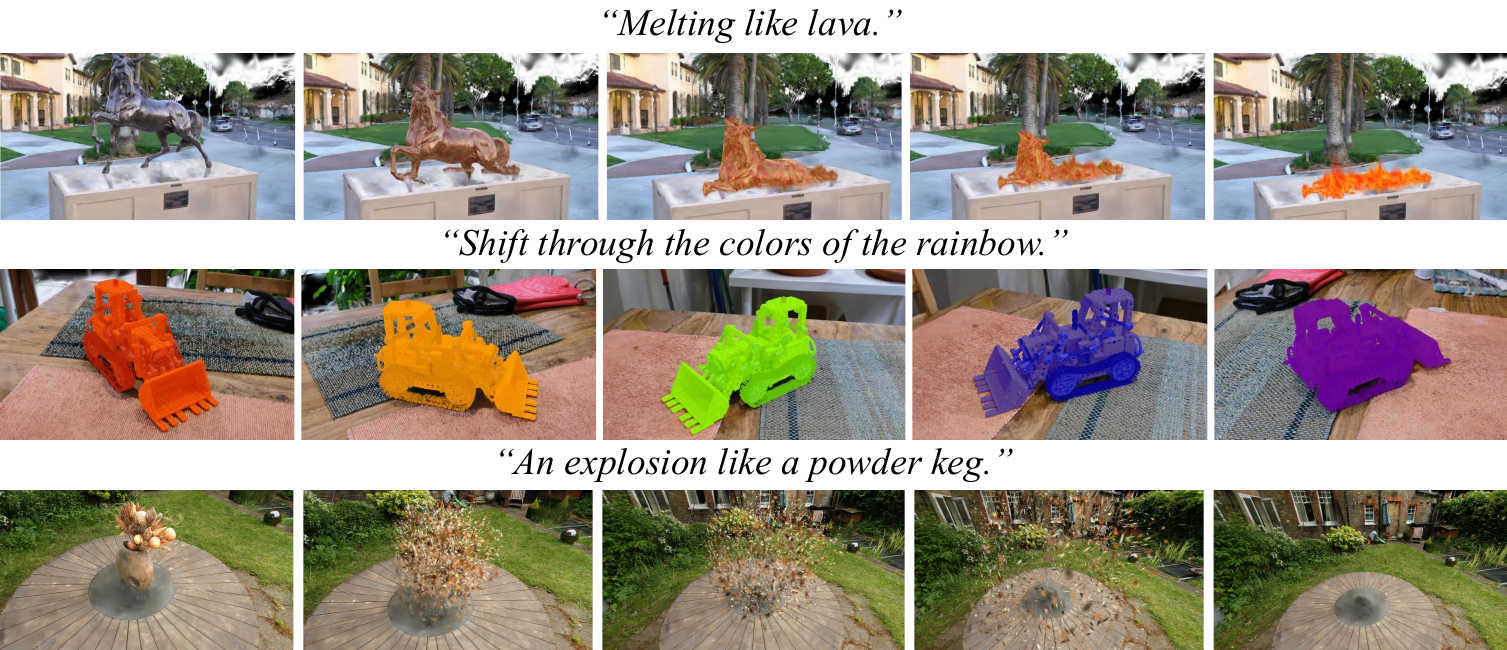}
    \vspace{-20pt}
    \caption{{\bf Qualitative results} showcasing the diversity and fidelity of animations generated by PromptVFX across different scenes and user prompts. Exact  prompts used to generate these animations are provided in the supplementary material.}
    \label{fig:qualitative}
\end{figure*}

\subsection{Time-Dependent Fields for Animation}
\label{subsec:time_dependent}
Rather than generating new geometry or using a physics simulator, we define animation through a continuous \emph{time-varying} field $f$ operating on each Gaussian attribute:
\begin{align}
    &f : \mathbb{R}^3 \times \mathbb{R}^3 \times \mathbb{R} \times [0,T] \mapsto \mathbb{R}^3 \times \mathbb{R}^3 \times \mathbb{R}\nonumber\\
    &f\left( \text{\textbf{A}}_i,t \right)
    = \left( f^\mu \left( \mu_i,t \right), f^\mathbf{c} \left( \mathbf{c}_i,t \right), f^\alpha \left( \alpha_i,t \right) \right)\label{eq:time_functions}\\
    &\qquad = \left( \mu_i(t), \mathbf{c}_i(t), \alpha_i(t) \right)
    =: \text{\textbf{A}}_i(t)\nonumber
\end{align}
where $i$ indicates a particular Gaussian, and $t \in [0, T], T \in \mathbb{R}$ spans the animation’s duration. These fields allow structured transformations such as translation, color transitions, or opacity changes over time.

\subsection{LLM-Based Translation of Text Instructions}
\label{subsec:llm_parsing}
We now describe how open-domain text instructions are converted into parametric fields that update each Gaussian’s center, color, and opacity. Our pipeline comprises multiple stages, each realized by a specialized LLM prompt. If additional renderings of the object are available, we can also involve a VLM for more precise attribute grounding. The detailed text instructions to the L(V)LMs that are mentioned here schematically can be found in the supplementary material.

\subsubsection{Design Phase}
A user provides an animation description (e.g., “move the vase up for two seconds, then dissolve it over one second”). We first request that the LLM break down this abstract instruction into animation phases, specifying approximate timings and actions. For instance, it might produce:
\begin{itemize}[leftmargin=1.25em]
    \item \texttt{Phase 1 (0-2s)}: “translate vase upward,”
    \item \texttt{Phase 2 (2-3s)}: “fade from opaque to transparent.”
\end{itemize}
In doing so, the LLM interprets the user’s high-level goals into a structured set of phases. If images of the object are provided, it can tailor the breakdown accordingly (e.g., factoring in bounding-box dimensions or shape cues). Simple static text instructions guarantee a structured output.

\subsubsection{Field Generation Phase}
The output from the desing phase is now translated by LLM calls into $\mu_i(t), \mathbf{c}_i(t), \alpha_i(t)$ (cf. Eq.~\ref{eq:time_functions}), still considering user-specified durations and transformations. For example, the geometry function might linearly interpolate the vase’s $z$-coordinate from $z_0$ to $z_0 + 2$ over $t \in [0, 2]$, while the opacity function steadily decreases $\alpha$ from 1 to 0 during $t \in [2,3]$.

\subsubsection{Animation Hypotheses Generation \& Scoring}
Some textual instructions might be ambiguous (e.g., “make the vase shake randomly”). For those, we can generate $m$ variations $f_1, f_2, \ldots, f_m$ of these parametric functions
\begin{align}
    f_j^\mu \left( \mu_i,t \right), f_j^\mathbf{c} \left( \mathbf{c}_i,t \right), f_j^\alpha \left( \alpha_i,t \right), \quad j \in \{ 1, 2, \ldots, m\}.
\end{align}
Each variation applies slightly different motion curves or color transitions. We then render each candidate animation (producing, for instance, a short sequence of frames) and feed these frames, along with the original text prompt, to a VLM that scores how well each animation matches the intended description on a $0$--$100$ scale. We select the highest-scoring candidate as our base animation.
We also include two optional refinement processes to improve the output.

\begin{figure*}[t]
    \centering
    \includegraphics[width=\linewidth]{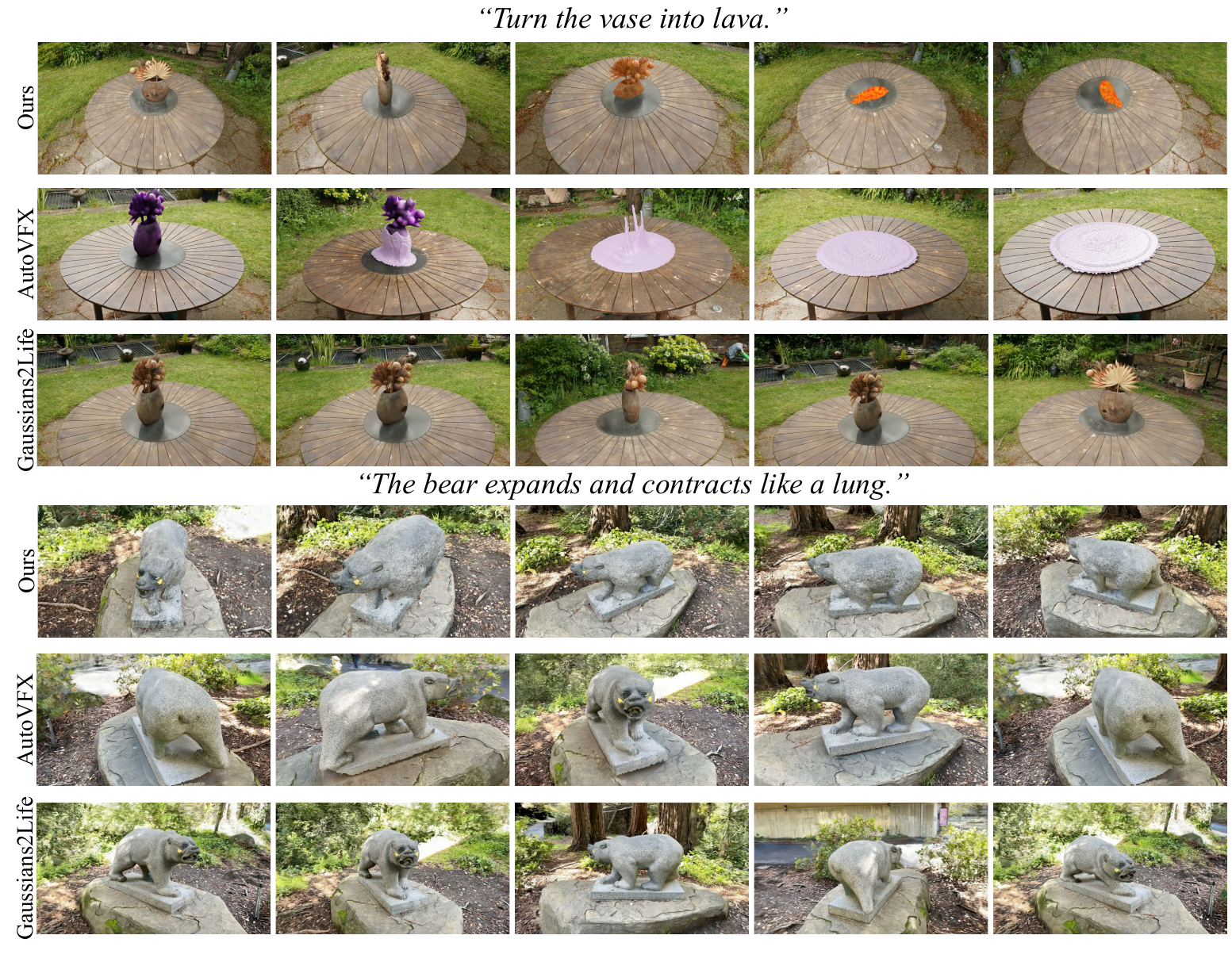}
    \vspace{-20pt}
    \caption{{\bf Qualitative comparison} with baselines on different scenes and user prompts. Our method achieves high-fidelity visual transformations and realistic motion, outperforming AutoVFX and Gaussians2Life. Exact prompts and additional qualitative comparisons are provided in the supplementary material.} 
    \label{fig:qualitative_comparison}
\end{figure*}

\begin{figure*}[t]
        \centering
        \includegraphics[width=1\linewidth]{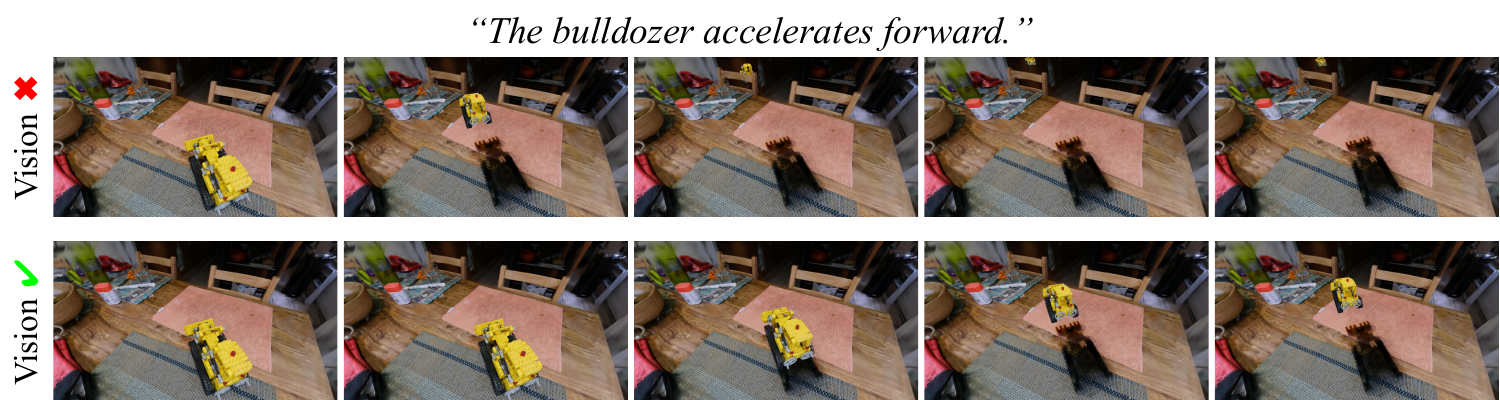}
        \vspace{-20pt}
        \caption{{\bf Impact of VLM} feedback on animation accuracy. Without visual feedback, the bulldozer incorrectly accelerates off the table, failing to account for scene constraints. With VLM refinement, the motion is corrected to remain contextually appropriate.}
        \label{fig:ablation-vision}
    \end{figure*}

\begin{figure*}[t]
        \centering
        \includegraphics[width=1\linewidth]{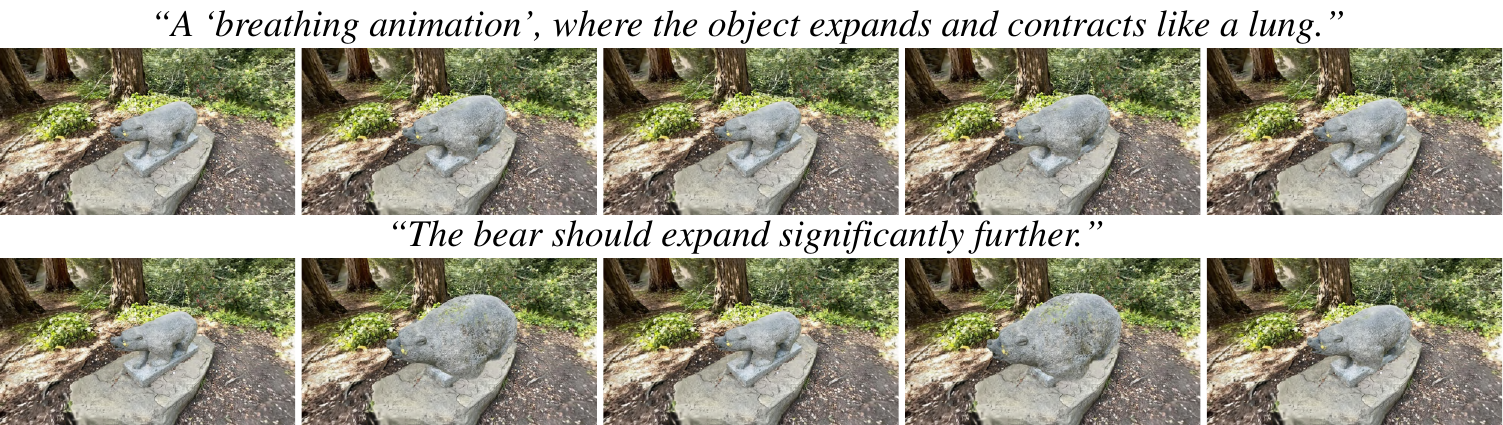}
        \vspace{-20pt}
        \caption{{\bf Animation refinement} is demonstrated through iterative user interaction. After generating an initial animation, the system enables users to provide follow-up prompts for further adjustments, enhancing control and precision in the final animation.}
        \label{fig:ablation-refinement}
    \end{figure*}

\begin{table*}[t]
  \centering
  \caption{
    Comparison of methods across CLIP similarity, VQAScore, and User Study ratings. 
    User Study includes Text Alignment and Animation Quality scores (1–5 Likert scale, normalized). 
    Bold indicates best per prompt and metric.
  }
  \label{tab:method-comparison}
  \vspace{-10pt}
  \resizebox{2\columnwidth}{!}{%
    \begin{tabular}{l l c c cc}
      \toprule
      \textbf{Prompt} & \textbf{Method} 
      & \textbf{CLIP~\cite{radford2021learning}} 
      & \textbf{VQAScore~\cite{lin2024evaluating}} 
      & \multicolumn{2}{c}{\textbf{User Study}} \\
      \cmidrule(lr){5-6}
      & 
      & 
      & 
      & \textbf{Text Alignment} 
      & \textbf{Animation Quality} \\
      \midrule
      \multirow{3}{*}{Turn the vase into lava.}
        & Gaussians2Life~\cite{wimmer2025gaussianstolife} & \textbf{0.206} & 0.196 & 0.043 & 0.235 \\ 
        & AutoVFX~\cite{hsu2024autovfx}                  & 0.159 & 0.444 & 0.615 & 0.485 \\ 
        & Ours                                           & 0.171 & \textbf{0.715} & \textbf{0.750} & \textbf{0.593} \\
      \midrule
      \multirow{3}{*}{The bear expands and contracts like a lung.}
        & Gaussians2Life~\cite{wimmer2025gaussianstolife} & 0.238 & 0.258 & 0.235 & 0.285 \\ 
        & AutoVFX~\cite{hsu2024autovfx}                  & \textbf{0.239} & \textbf{0.283} & 0.073 & 0.315 \\ 
        & Ours                                           & 0.204 & \textbf{0.283} & \textbf{0.693} & \textbf{0.465} \\ 
      \midrule
      \multirow{3}{*}{The bulldozer accelerates forward.}
        & Gaussians2Life~\cite{wimmer2025gaussianstolife} & \textbf{0.308} & 0.681 & 0.278 & 0.235 \\ 
        & AutoVFX~\cite{hsu2024autovfx}                  & 0.172 & 0.451 & 0.035 & 0.058 \\ 
        & Ours                                           & 0.227 & \textbf{0.685} & \textbf{0.893} & \textbf{0.685} \\ 
      \bottomrule
    \end{tabular}
  }
\end{table*}

\begin{figure*}[t]
        \centering
        \includegraphics[width=1\linewidth]{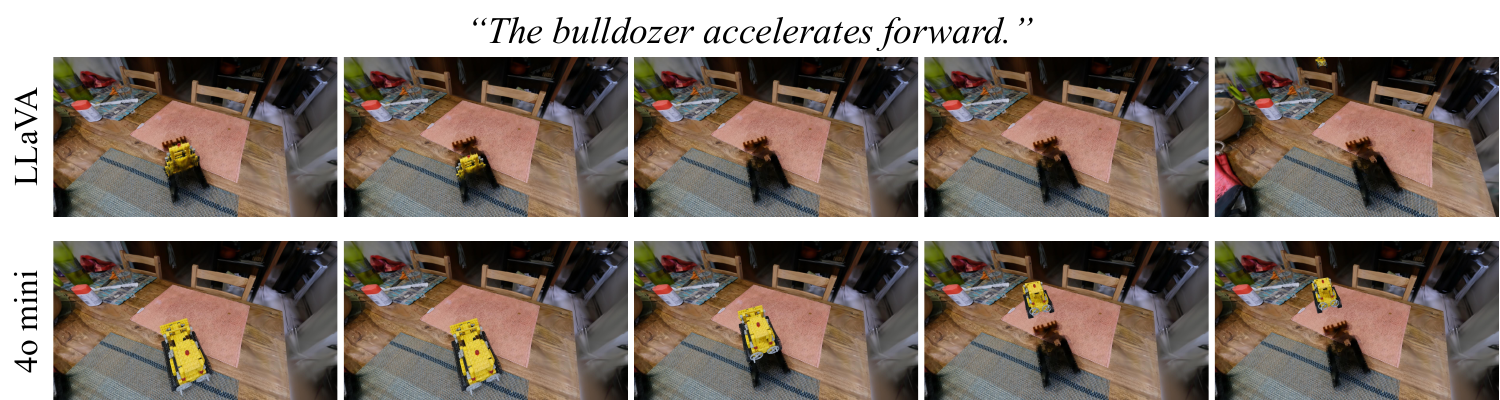}
        \vspace{-20pt}
        \caption{{\bf VLM comparison} for animation quality effect with different vision-language models (VLMs) that interpret the prompt. Given the same input text, GPT-4o-mini produces a coherent bulldozer acceleration, while LLaVA struggles to maintain object consistency and motion realism.}
        \label{fig:ablation-llava}
    \end{figure*}

\subsubsection{Automatic Refinement}
To further refine the chosen base animation, we prompt a specialized VLM that compares the user’s description and the rendered frames. It identifies discrepancies (e.g., “the vase does not rise high enough”) and adjusts our field functions $\mu_i(t), \mathbf{c}_i(t), \alpha_i(t)$ to more closely match the user’s instructions.

\subsubsection{Conversational Refinement}
If a user observes misinterpretations of the intended animation, open-text feedback can be provided (e.g., “spin faster in the first second, fade more quickly in the final half-second”). We combine this feedback with the existing field $A_i$ and animation frames, asking the VLM to produce another iteration of refined functions. This conversational loop can continue until the user is satisfied. This enables rapid iteration cycles with a natural text-driven interface due to iteration without computational overhead from diffusion models or physics-based simulations.
\section{Experiments}
\label{sec:experiments}

\subsection{Experimental Details}

\textbf{Dataset \& Preprocessing.}
We evaluate our method on real-world scenes from Mip-NeRF360~\cite{barron2022mipnerf360} (the \textit{garden vase} and \textit{bulldozer}), \textit{bear} scene from Instruct-NeRF2NeRF~\cite{instructnerf2023} and an additional \textit{horse} scene from Tanks and Temples~\cite{Knapitsch2017}.
Each scene is reconstructed via a Gaussian splatting pipeline, providing a set of Gaussians encoding geometry and color. 
When animating, user instructions (e.g., “the vase”) are mapped to corresponding Gaussians.

\textbf{Baselines.}
We compare against Gaussians2Life~\cite{wimmer2025gaussianstolife}, which uses video diffusion to synthesize 2D motion lifted to 3D Gaussians via optimization, and AutoVFX~\cite{hsu2024autovfx}, which generates physically simulated animations through Blender scripts, requiring mesh extraction and offline rendering. Both pipelines depend on heavy preprocessing and lack interactivity. In contrast, our method applies functional transformations directly to Gaussians, enabling real-time animation without diffusion or physics simulation.

\textbf{Implementation Notes.}
Our method is implemented in Python and leverages GPT-4~\cite{achiam2023gpt} to parse text prompts and generate animation functions. All experiments are conducted on a single NVIDIA RTX 4090 GPU.

\subsection{Qualitative Evaluation}
We demonstrate the capabilities of PromptVFX across a variety of scenes and user instructions.  
\autoref{fig:qualitative} shows that our system can generate animations involving motion, color changes, and opacity variations in response to natural language prompts.  
Each animation is produced by updating Gaussian parameters such as position and appearance in real time, enabling interactive feedback and rapid iteration during the creative process.

\textbf{Comparison with Baselines.}  
\autoref{fig:qualitative_comparison} provides a visual comparison between PromptVFX and two recent methods, Gaussians2Life~\cite{wimmer2025gaussianstolife} and AutoVFX~\cite{hsu2024autovfx}. Unlike these approaches, which rely on diffusion models or physics simulations and require offline processing, our framework applies direct functional transformations to the Gaussians.  
This design results in more responsive editing, better temporal coherence, and consistent appearance transformations across views. PromptVFX delivers high-quality animations that align with user prompts while maintaining the speed and flexibility required for interactive content creation.

\subsection{Quantitative Evaluation}

Evaluating 3D animation quality is difficult due to the absence of ground-truth sequences. Following prior work~\cite{chen2024dge,hsu2024autovfx,wimmer2025gaussianstolife}, we report CLIP similarity~\cite{radford2021learning} as a frame-level proxy for text-to-image alignment. However, CLIP does not capture motion or temporal coherence. We therefore also report VQAScore, a video-level metric shown to better align with human judgment~\cite{lin2024evaluating}. It estimates the probability that a model answers “Yes” to the prompt: \emph{Does this video align with the described animation: ``\{prompt\}''?}, denoted as $\mathbb{P}(\text{``Yes''} \mid \text{video, prompt})$.

Table~\ref{tab:method-comparison} shows that \textbf{PromptVFX} achieves the highest VQAScore~\cite{lin2024evaluating} across all evaluated prompts, demonstrating better text-video alignment from a temporal and semantic perspective. While Gaussians2Life~\cite{wimmer2025gaussianstolife} ranks higher in CLIP~\cite{radford2021learning} for some cases due to preserving frame-level appearance, it does not always reflect coherent or realistic motion.

To further evaluate perceptual quality, we conducted a user study with 35 participants, following the protocol from AutoVFX~\cite{hsu2024autovfx}. Participants rated the generated videos on two criteria: \textit{Text Alignment} and \textit{Animation Quality}, using a 1-5 Likert scale (normalized to [0, 1] using min-max normalization). As shown in Table~\ref{tab:method-comparison}, PromptVFX outperforms all baselines across both criteria and all prompts, demonstrating its effectiveness in producing coherent and semantically grounded 3D animations.

\subsection{Ablation Studies}
We conduct several ablation studies to evaluate the impact of different components in our pipeline.

\vspace{-10pt}
\subsubsection{Impact of Vision-Language Model Feedback}
To assess the role of VLM-based feedback, we compare animations generated with and without visual refinement. As shown in \autoref{fig:ablation-vision}, without visual feedback, the bulldozer accelerates off the table, failing to respect scene constraints. When rendered frames are provided to the VLM for refinement, the animation is corrected to maintain a physically plausible motion. This highlights the importance of incorporating visual guidance in aligning animations with scene context.

\vspace{-10pt}
\subsubsection{User-Guided Animation Refinement}
We examine the benefits of allowing iterative user feedback in refining animations. In \autoref{fig:ablation-refinement} an initial animation is generated based on the text prompt, after which the user provides a follow-up adjustment. The refined animation better captures the intended motion, demonstrating how interactive refinements improve animation precision. This experiment underscores the importance of keeping the user in the loop for fine-grained control over the animation generation process.

\vspace{-10pt}
\subsubsection{Effect of VLM Choice on Animation Quality}
We compare different vision-language models used in our pipeline. \autoref{fig:ablation-llava} illustrates that GPT-4o-mini produces a fluent bulldozer acceleration, while LLaVA~\cite{liu2023llava, liu2024llavanext} struggles with maintaining object consistency and motion realism.

This suggests that model selection significantly affects animation quality and highlights the need for well-designed system prompts tailored to the VLM’s capabilities.

    \begin{figure}[t]
        \centering
        \includegraphics[width=1\linewidth]{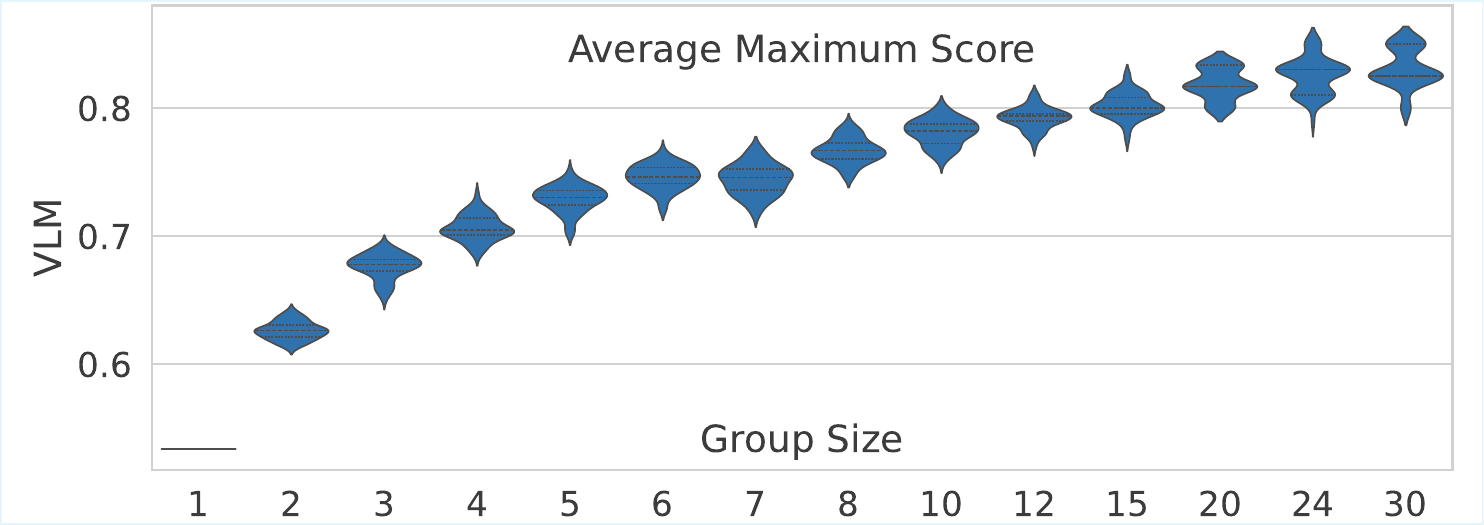}
        \vspace{-20pt}
        \caption{{\bf Best Scores for various amount of hypotheses.} Impact of the number of hypotheses on VLM scores. As the group size increases, VLM scores improve.}
        \label{fig:ablation-hypotheses}
    \end{figure}

\vspace{-10pt}
\subsubsection{Effect of Hypothesis Sampling}
We analyze how the number of generated animation hypotheses influences performance. \autoref{fig:ablation-hypotheses} shows that increasing the number of candidate animations improves VLM scores. The VLM score is obtained by querying a vision-language model to evaluate how well snapshots of the animation from different viewpoints match the original text prompt, providing a more holistic assessment of animation coherence. However, we observe diminishing returns beyond a certain number of hypotheses, indicating that an optimal balance exists between diversity and computational efficiency.
\vspace{-5pt}
\section{Limitations}
\label{sec:limitations}
While PromptVFX delivers fast, interactive text-driven volumetric animations, it has several constraints. It cannot support collisions or contact interactions. It can only animate existing Gaussians, so particle effects such as smoke or fire cannot be generated. Without semantic scene understanding, relational prompts such as “place the vase under the table” may not be interpreted correctly. In future work, we plan to integrate physics priors, enable dynamic splat insertion for particle effects, and incorporate semantic object labels to address these limitations.

\vspace{-5pt}
\section{Conclusion}
\label{sec:conclusion}
This paper introduces PromptVFX, a framework that reformulates 3D animation as a field prediction task by applying time-varying transformations directly to Gaussian splats. Unlike diffusion-based or physics-driven pipelines, our method enables real-time, text-driven animation through parametric updates generated by large language and vision-language models. The approach reduces manual effort, supports fast iteration, and produces high-fidelity, temporally consistent results. By lowering the barriers to 3D animation, this work takes a step toward democratizing open-ended visual effects creation.
\vspace{-14pt}
{
    \small
    \vspace{-14pt}
    \bibliographystyle{ieeenat_fullname}
    \bibliography{main}
}
\clearpage
\setcounter{page}{1}
\appendix

\section{Ethical Considerations: Deep Fakes}
\label{sec:ethical_considerations}
As with many generative AI methods, PromptVFX introduces ethical concerns regarding the potential misuse of synthetic media for deceptive purposes. While our approach is intended to empower creativity and streamline 3D animation workflows, we acknowledge the broader implications associated with AI-generated content, especially where distinguishing between real and synthetic media is crucial. We emphasize the importance of responsible usage and advocate for generative animation as a means to support artistic expression and accessibility rather than manipulation or misinformation.

\section{Implementation Details}
In this section, we provide detailed descriptions of the prompt templates, specific animation prompts used in our experiments to support the reproducibility and transparency of our results.
%
%
\subsection{Prompt design}

\begin{figure}[!htbp]
    \centering
    \includegraphics[width=1\linewidth]{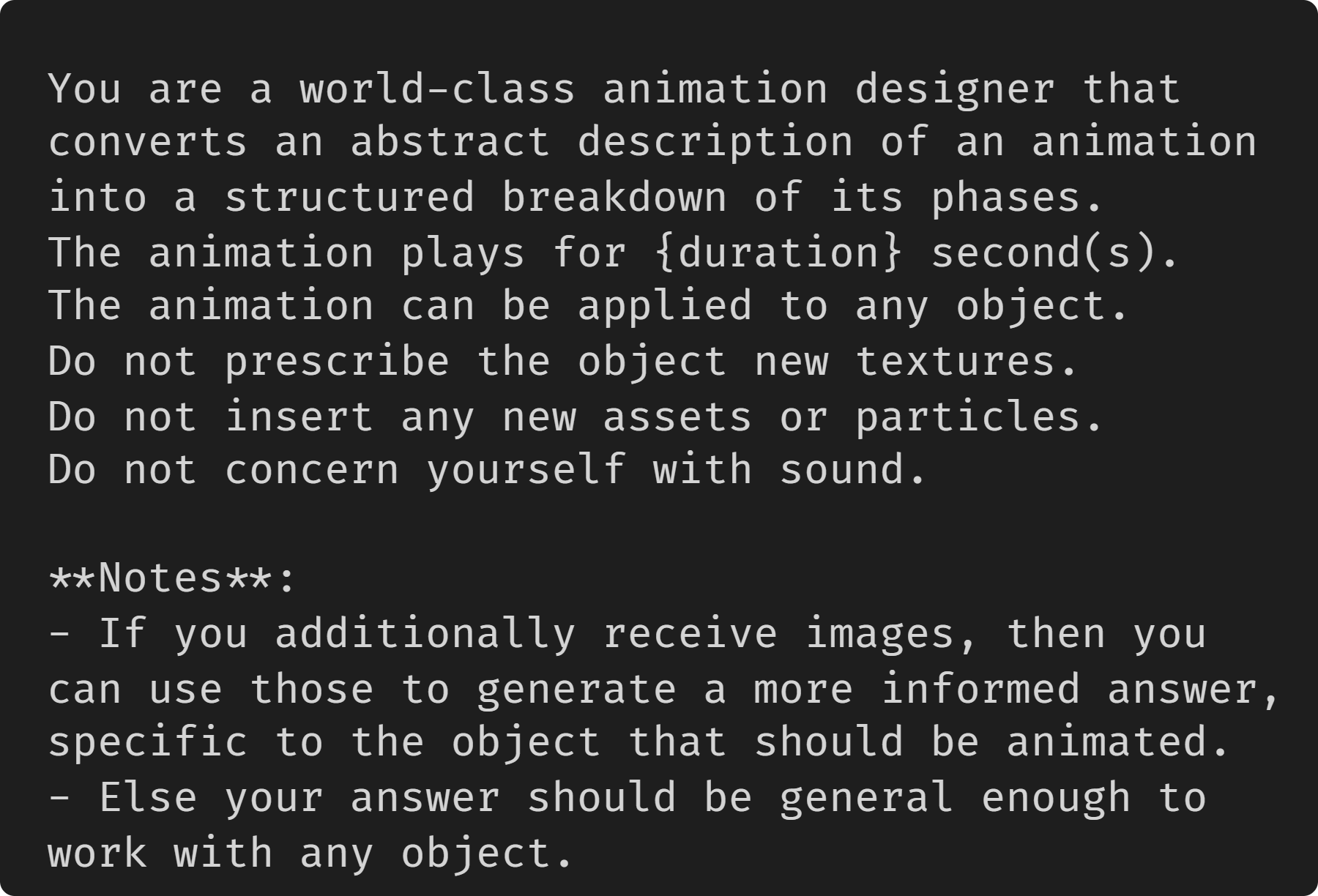}
    \caption{System Message for breaking down an abstract animation description into concrete animation phases.}
    \label{fig:abstract-summary}
\end{figure}

\begin{figure}
    \centering
    \includegraphics[width=1\linewidth]{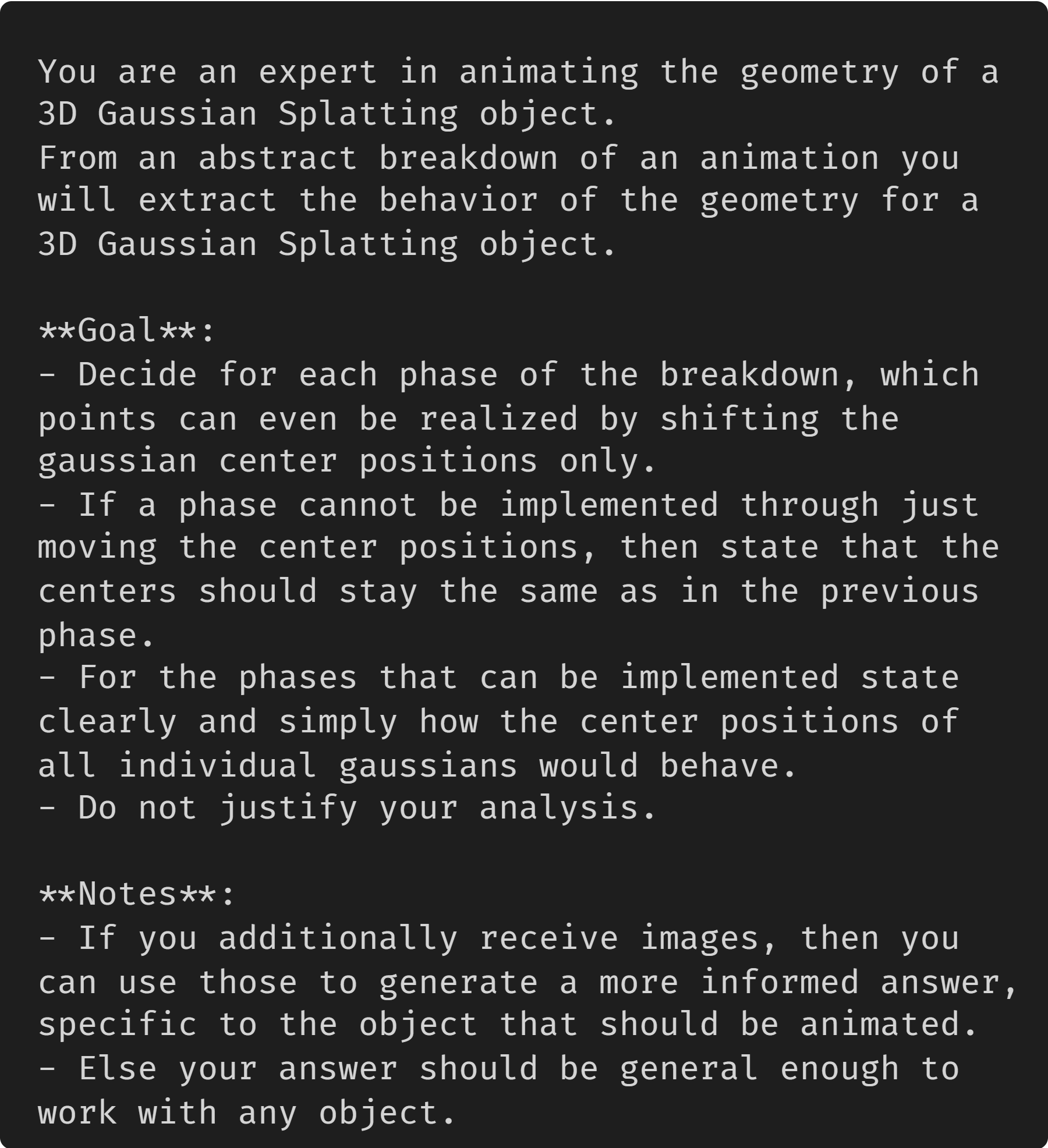}
    \caption{System Message for deriving the behavior of the Gaussian's center positions during the animation phases.}
    \label{fig:centers-behavior}
\end{figure}

\begin{figure}
    \centering
    \includegraphics[width=1\linewidth]{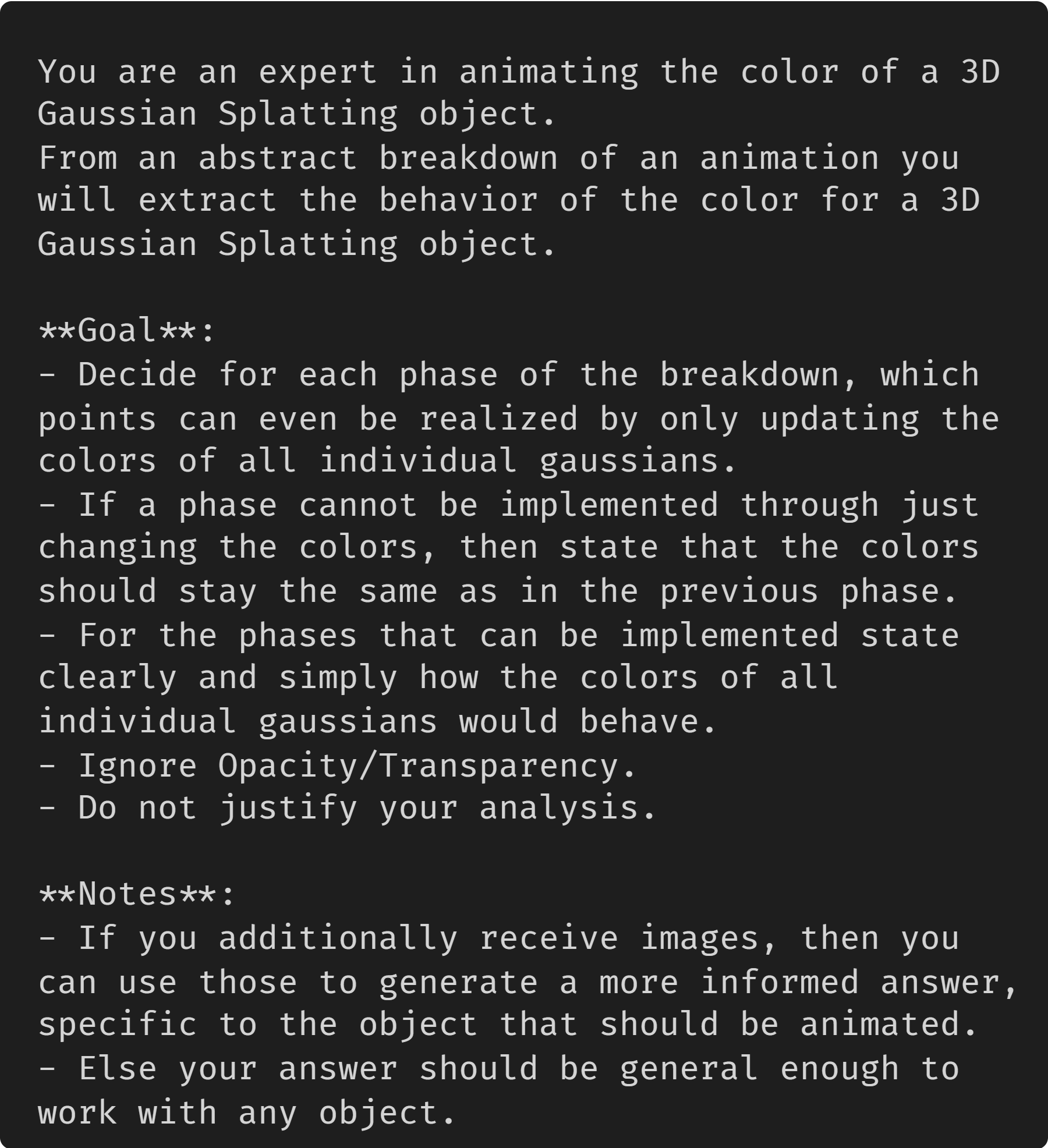}
    \caption{System Message for deriving the behavior of the Gaussian's colors during the animation phases.}
    \label{fig:rgbs-behavior}
\end{figure}

\begin{figure}
    \centering
    \includegraphics[width=1\linewidth]{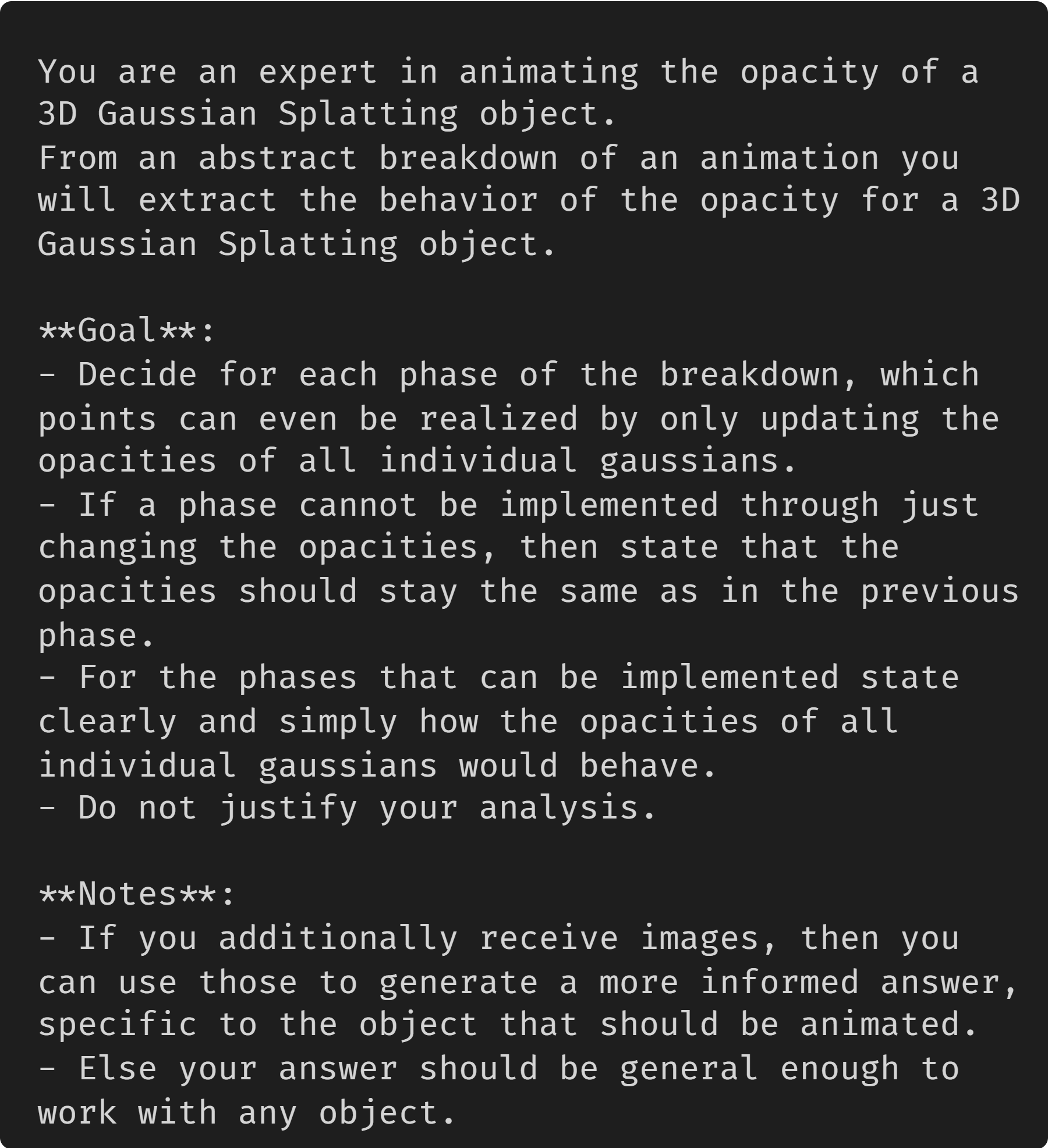}
    \caption{System Message for deriving the behavior of the Gaussian's opacity values during the animation phases.}
    \label{fig:opacities-behavior}
\end{figure}

\begin{figure}
    \centering
    \includegraphics[width=1\linewidth]{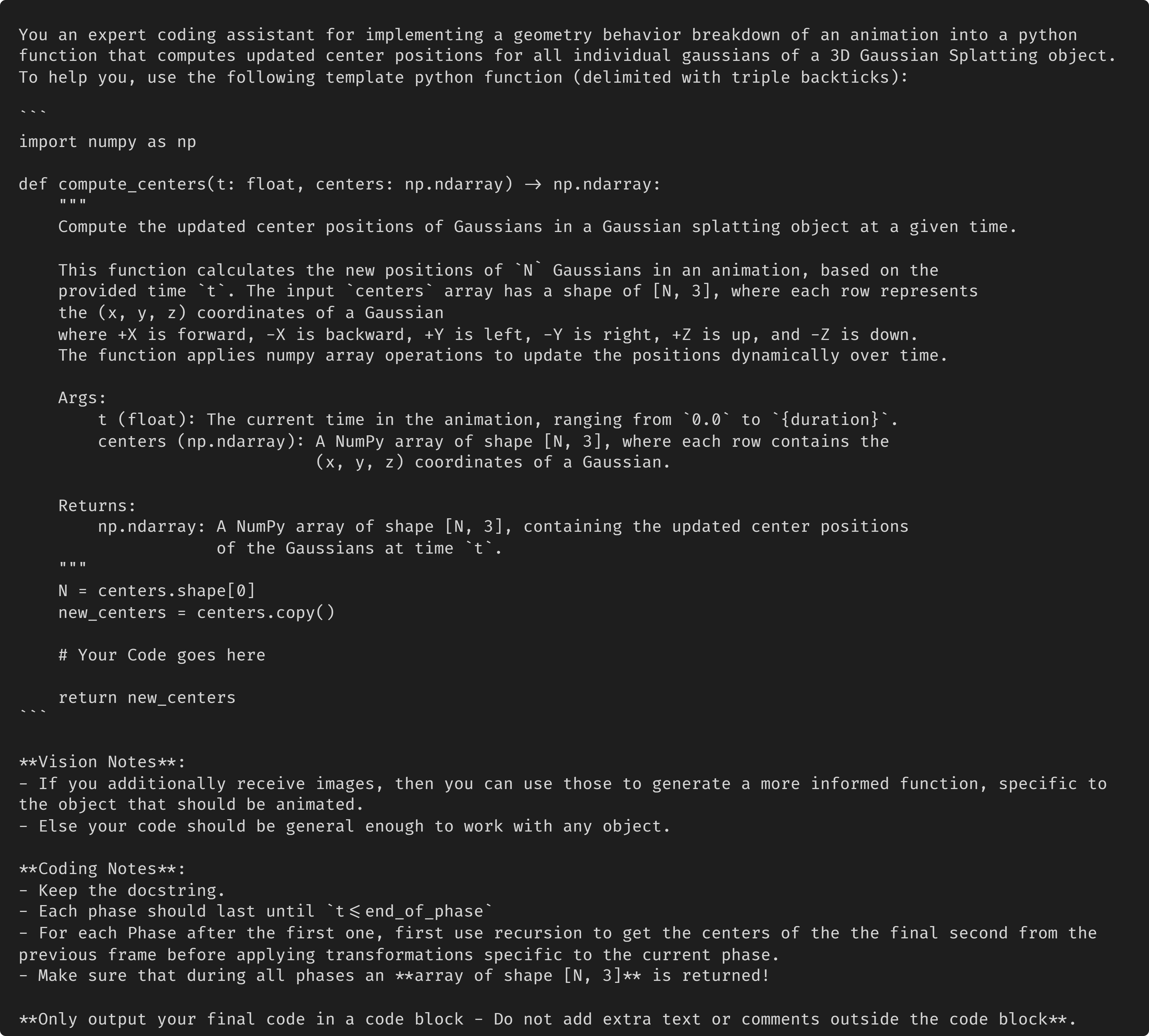}
    \caption{System message for generating Gaussian center position transformations.}
    \label{fig:code-centers}
\end{figure}

\begin{figure}
    \centering
    \includegraphics[width=1\linewidth]{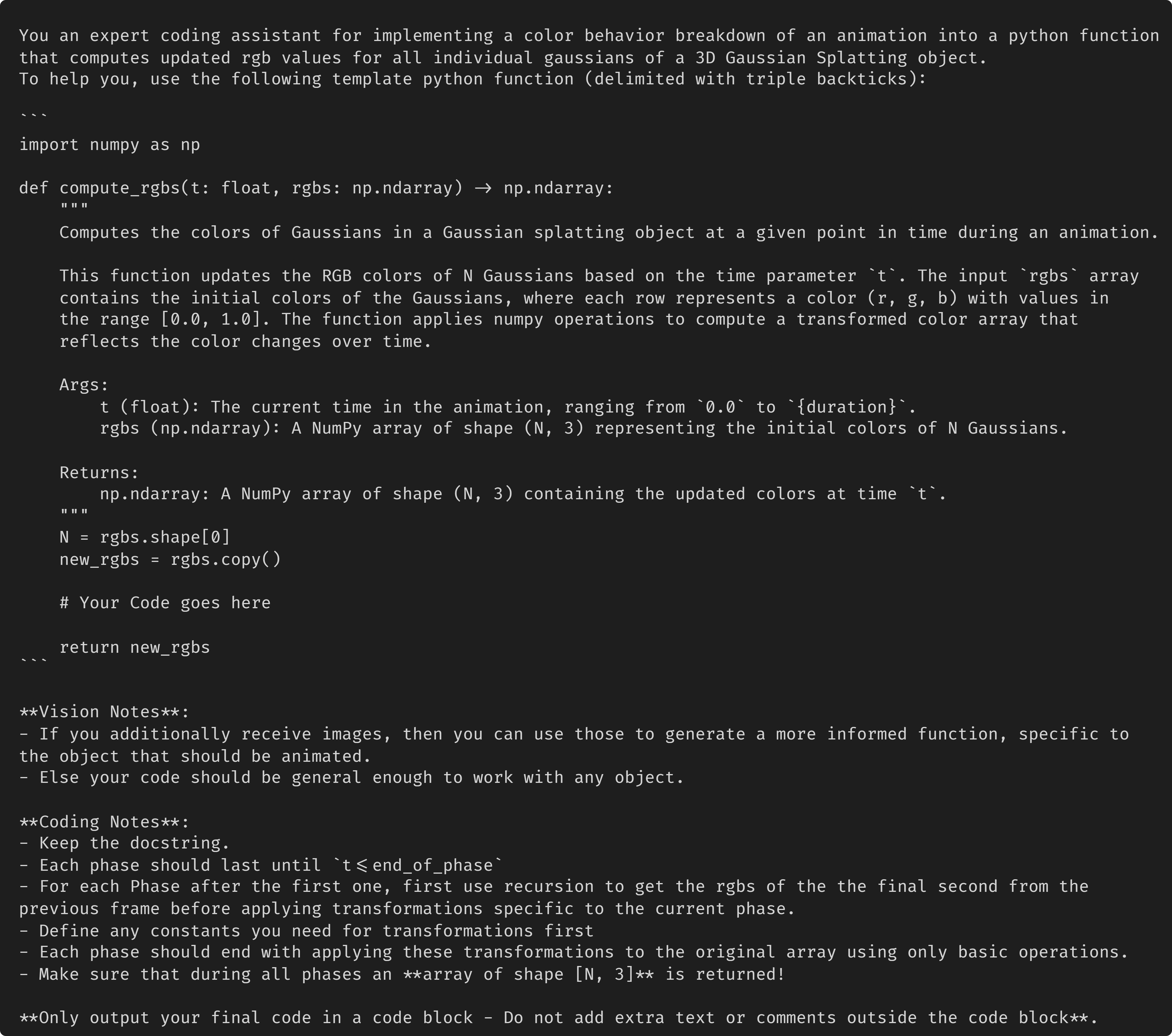}
    \caption{System message for generating Gaussian color transformations.}
    \label{fig:code-rgbs}
\end{figure}

\begin{figure*}
    \centering
    \includegraphics[width=1\linewidth]{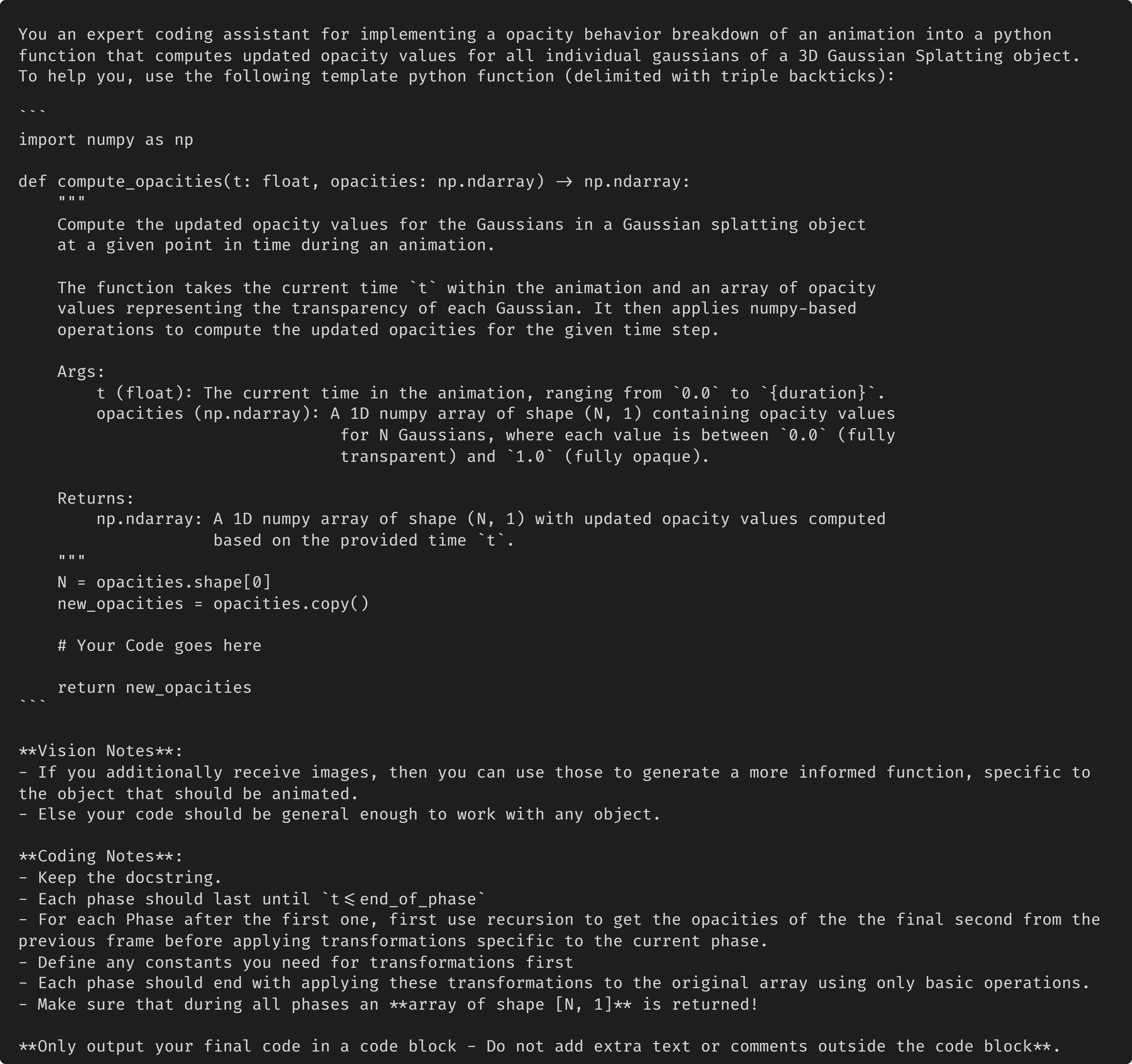}
    \caption{System message for generating Gaussian opacity transformations.}
    \label{fig:code-opacities}
\end{figure*}

\begin{figure}
    \centering
    \includegraphics[width=1\linewidth]{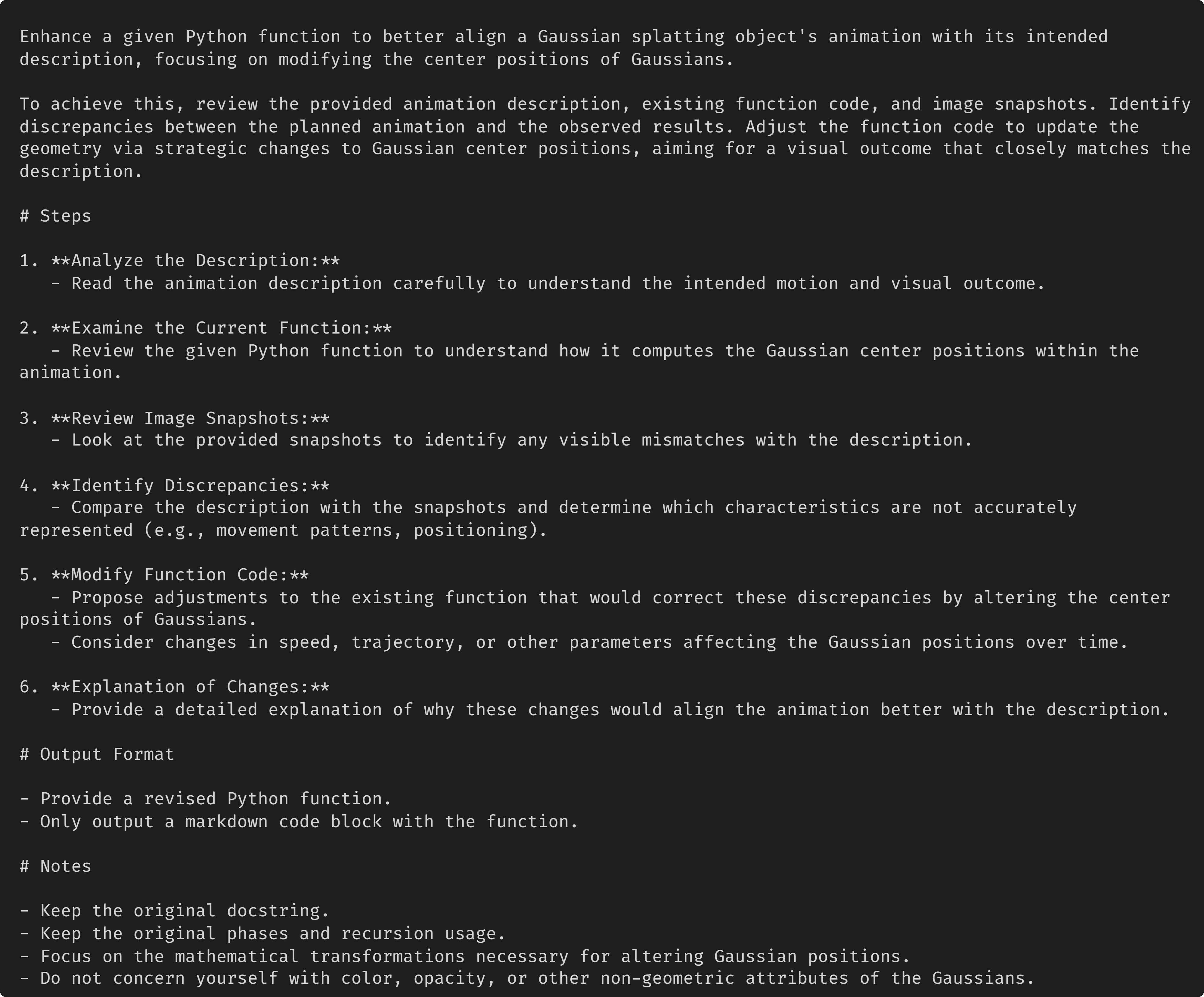}
    \caption{System message for automatically improving Gaussian center position transformations.}
    \label{fig:auto-improve-centers}
\end{figure}

\begin{figure}
    \centering
    \includegraphics[width=1\linewidth]{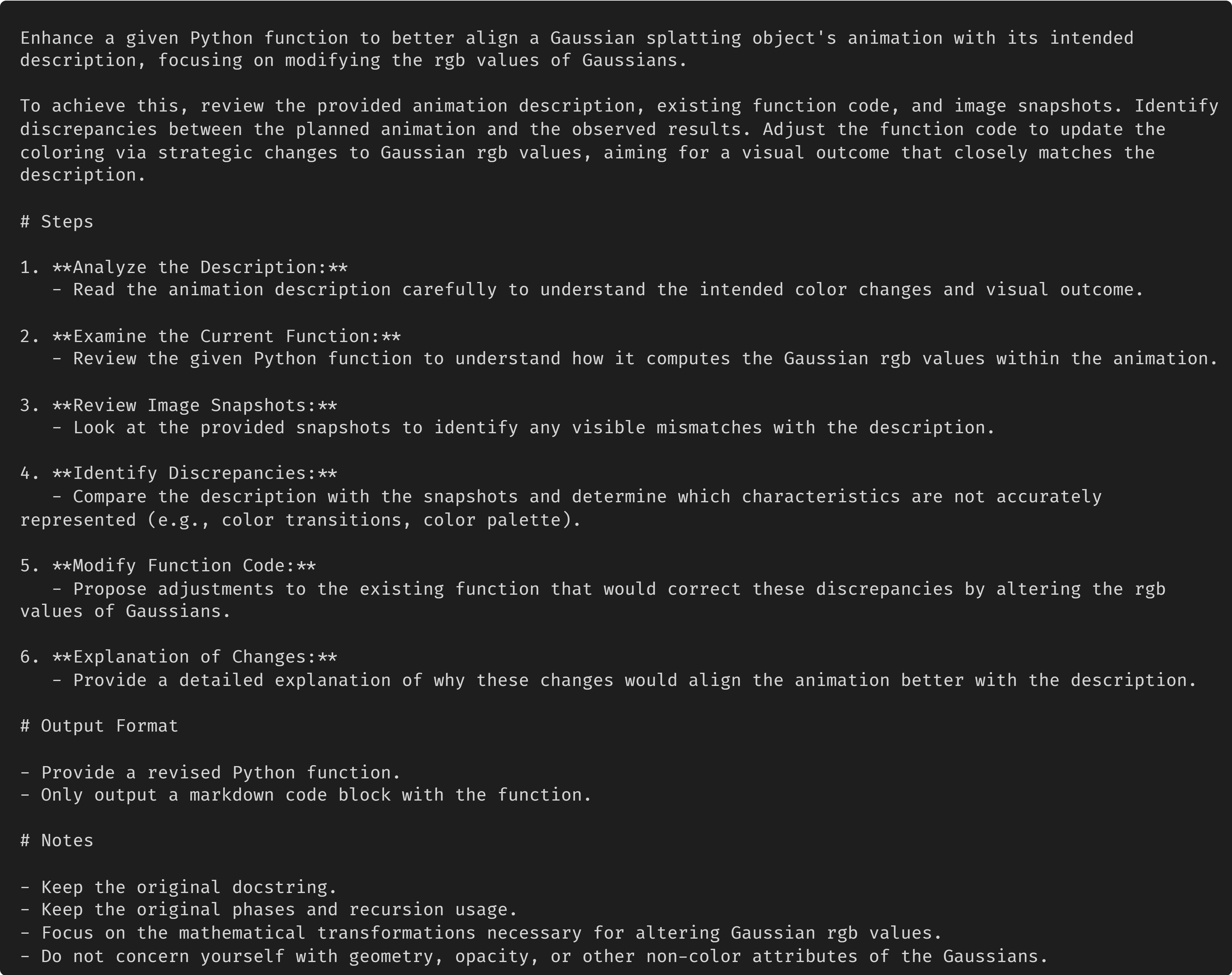}
    \caption{System message for automatically improving Gaussian color transformations.}
    \label{fig:auto-improve-rgbs}
\end{figure}

\begin{figure}
    \centering
    \includegraphics[width=1\linewidth]{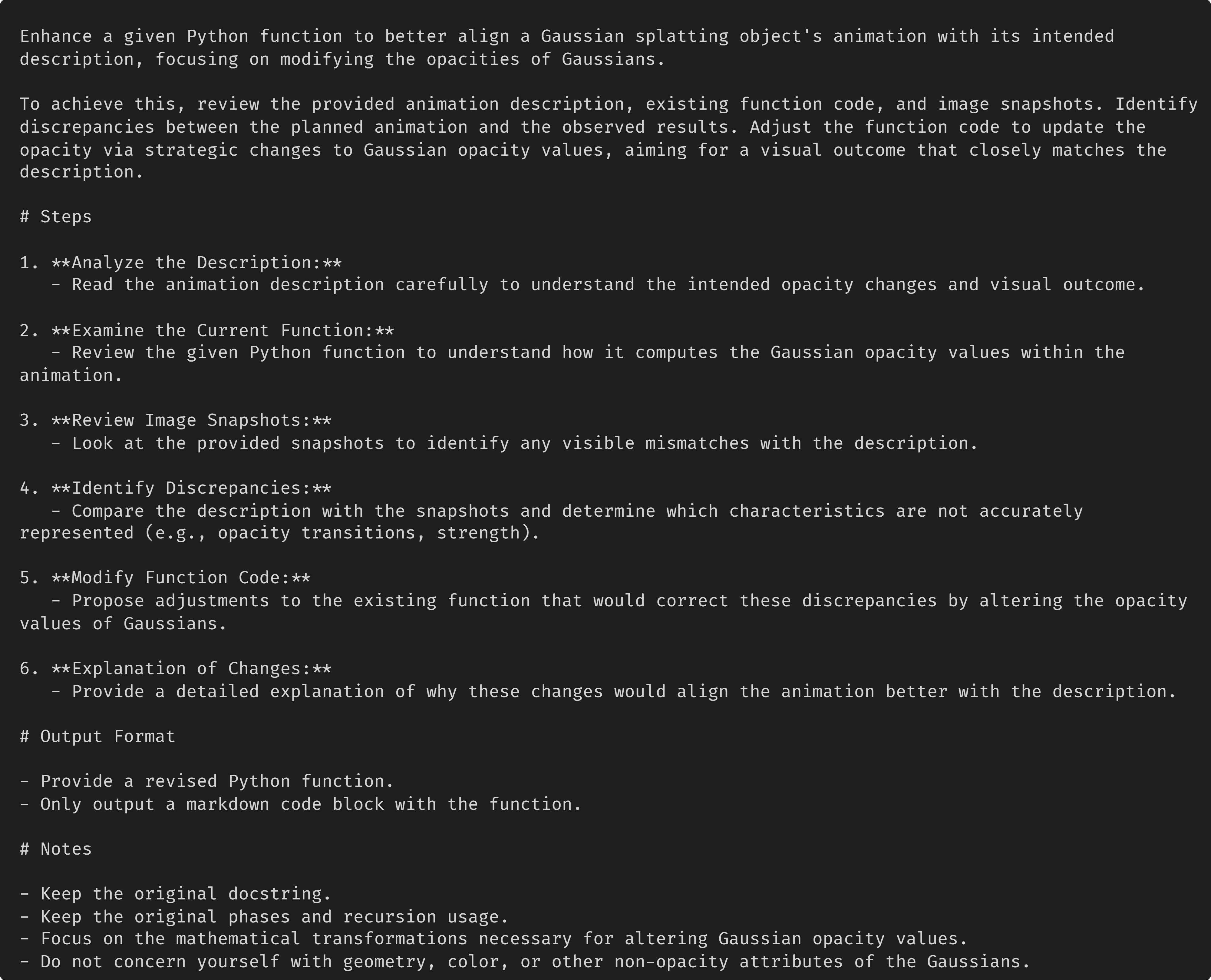}
    \caption{System message for automatically improving Gaussian opacity transformations.}
    \label{fig:auto-improve-opacities}
\end{figure}

\begin{figure}
    \centering
    \includegraphics[width=1\linewidth]{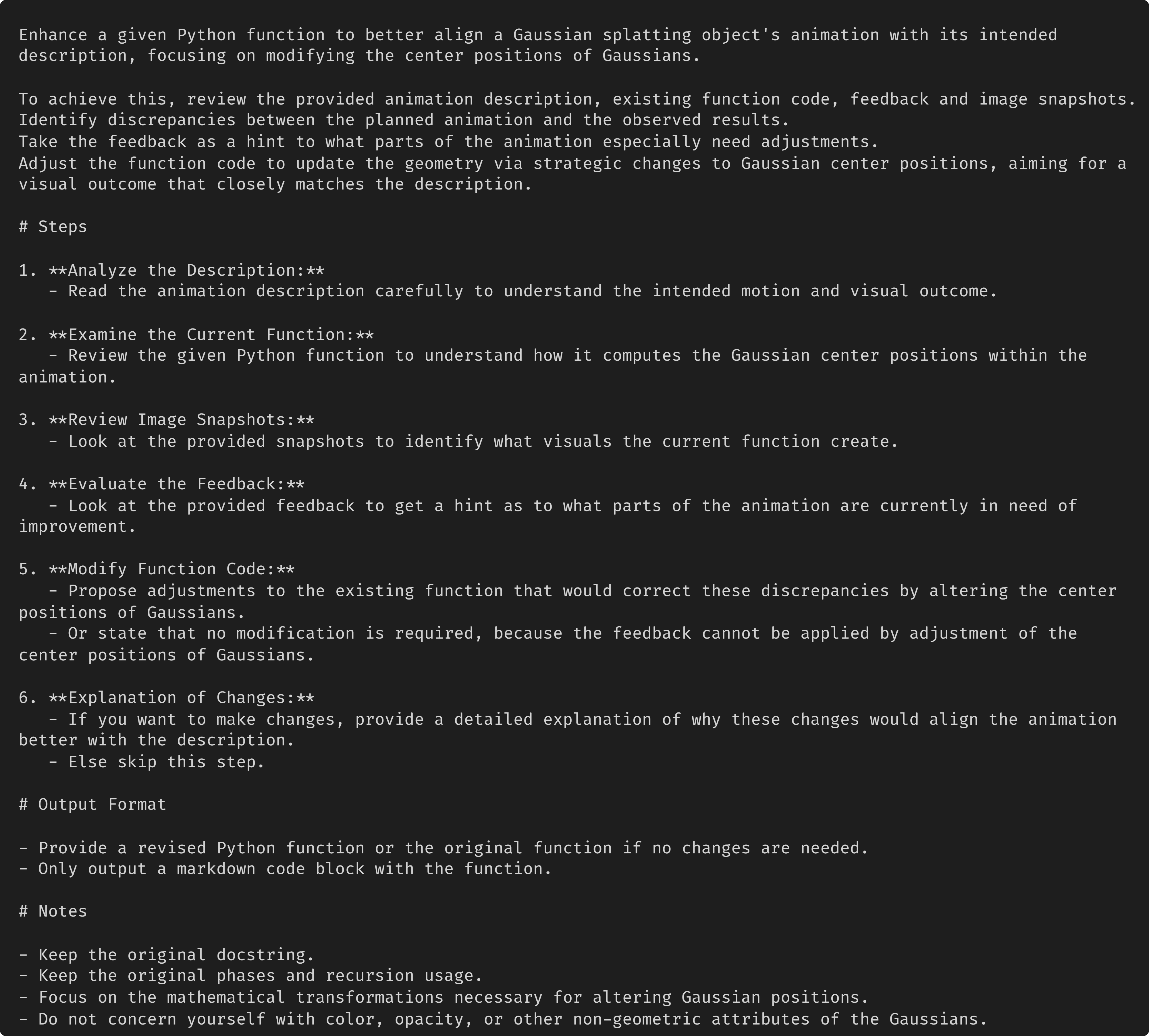}
    \caption{System message for refining Gaussian center transformations based on user feedback.}
    \label{fig:feedback-centers}
\end{figure}

\begin{figure}
    \centering
    \includegraphics[width=1\linewidth]{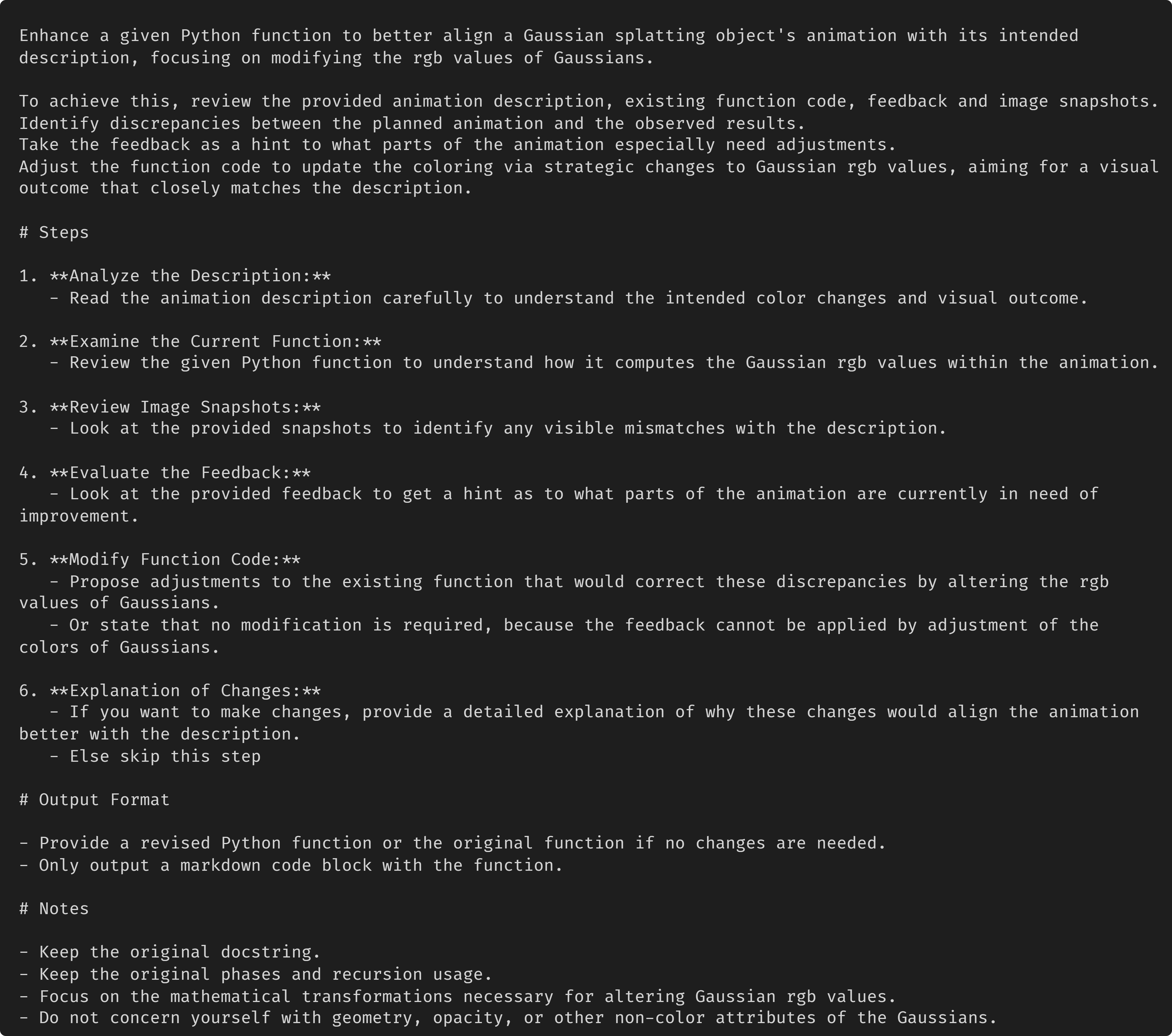}
    \caption{System message for refining Gaussian color transformations based on user feedback.}
    \label{fig:feedback-rgbs}
\end{figure}

\begin{figure}
    \centering
    \includegraphics[width=1\linewidth]{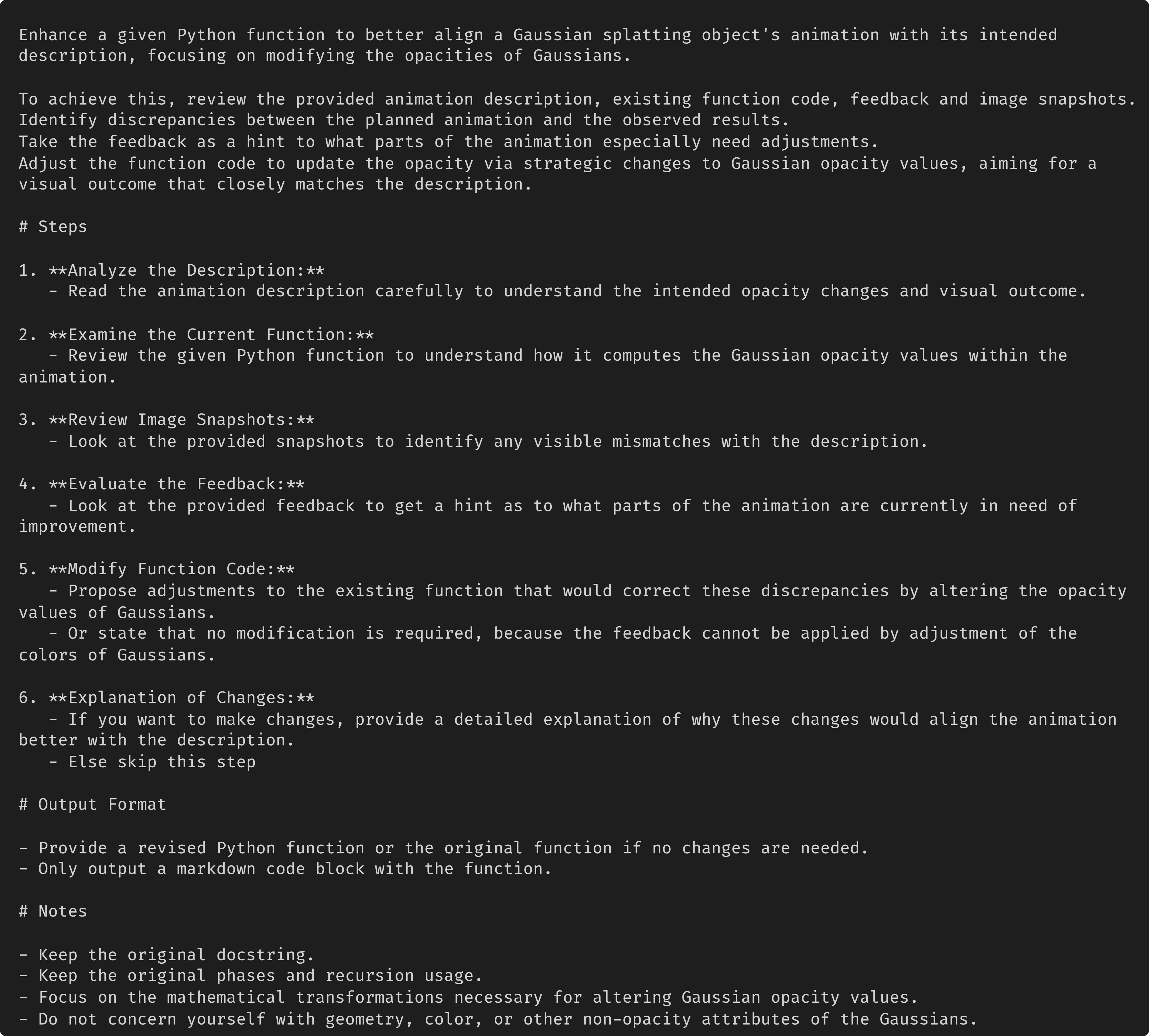}
    \caption{System message for refining Gaussian opacity transformations based on user feedback.}
    \label{fig:feedback-opacities}
\end{figure}

\begin{figure}
    \centering
    \includegraphics[width=0.8\linewidth]{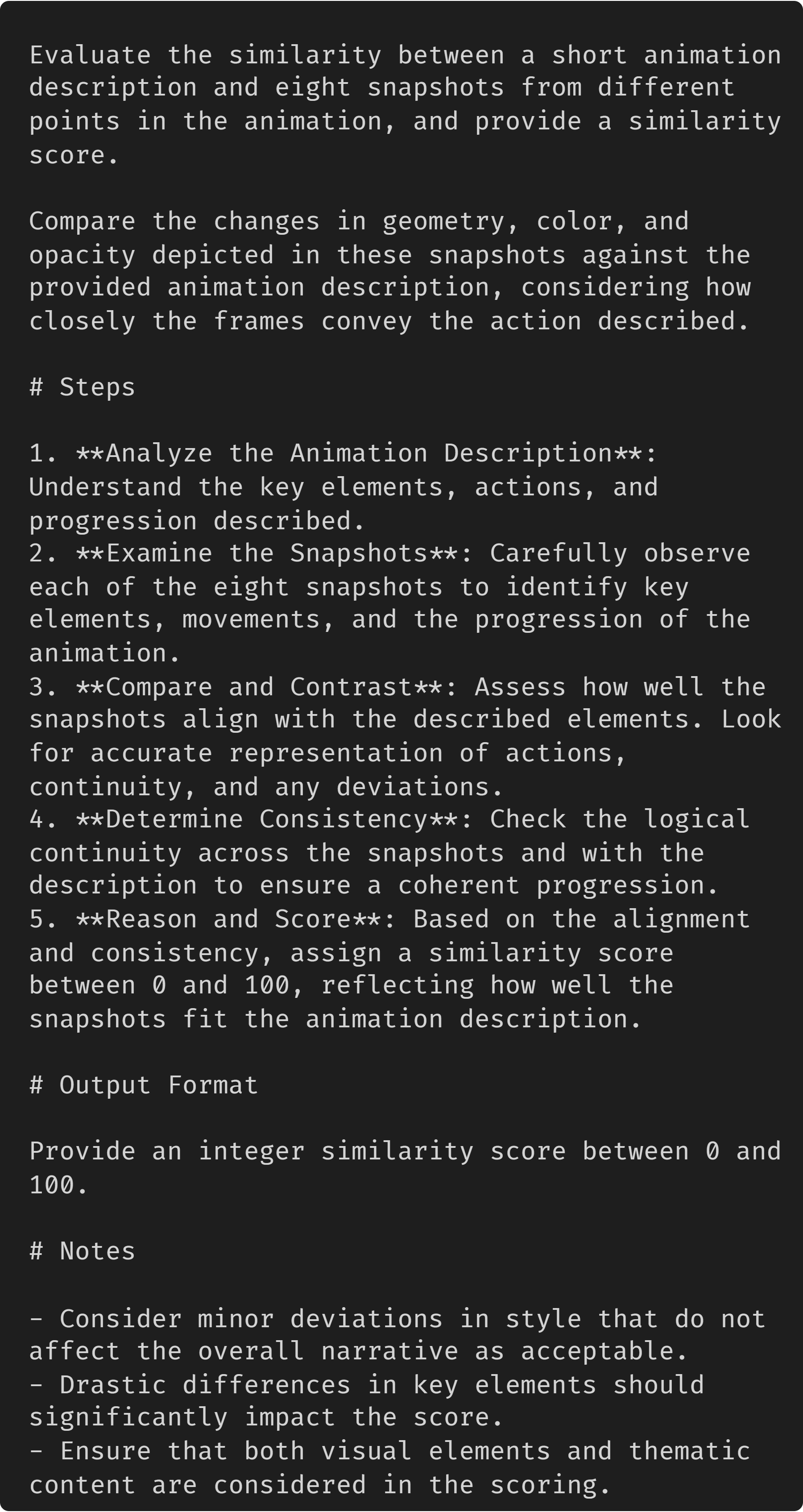}
    \caption{System message for evaluating the alignment of animations to their textual descriptions.}
    \label{fig:score-animation}
\end{figure}

We present the system prompts used in our pipeline for interacting with the L(V)LM to generate animations. Each prompt is designed to guide different stages of the animation process, from parsing abstract user instructions to generating and refining animation functions. \autoref{fig:abstract-summary} to \autoref{fig:score-animation} illustrate the various system prompts employed across different modules of our framework.

%
%
%
%
%
%
%
%

\begin{figure}
    \centering
    \includegraphics[width=1\linewidth]{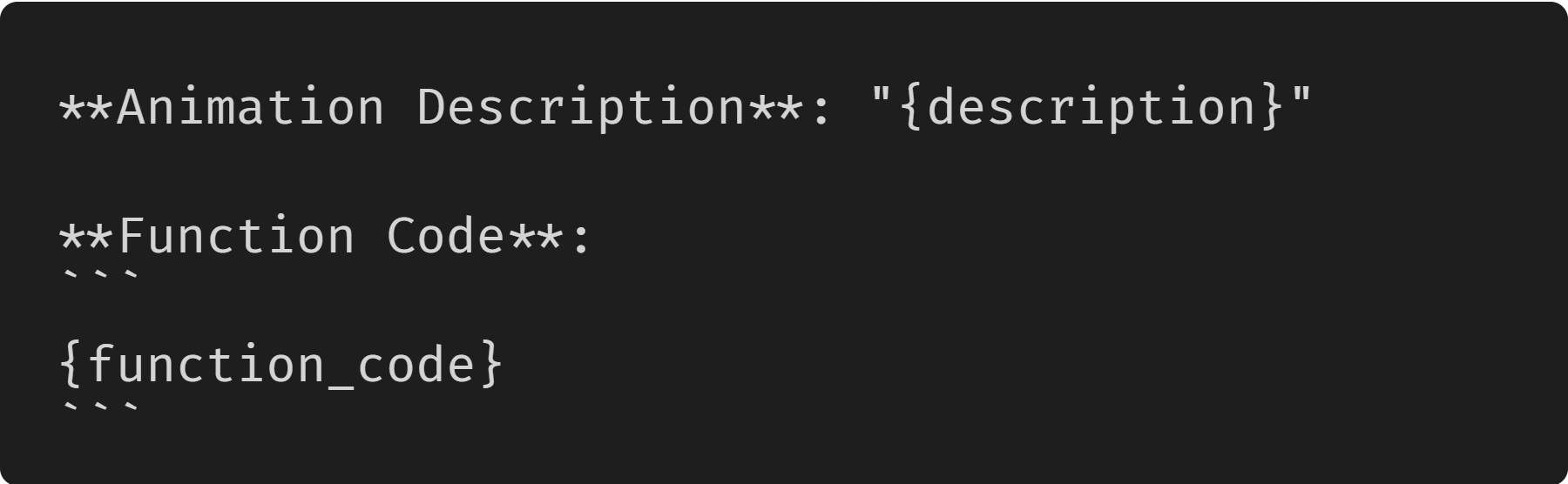}
    \caption{User message template for automatically improving Gaussian transformations.}
    \label{fig:auto-improve-user}
\end{figure}
%
%
\begin{figure}
    \centering
    \includegraphics[width=1\linewidth]{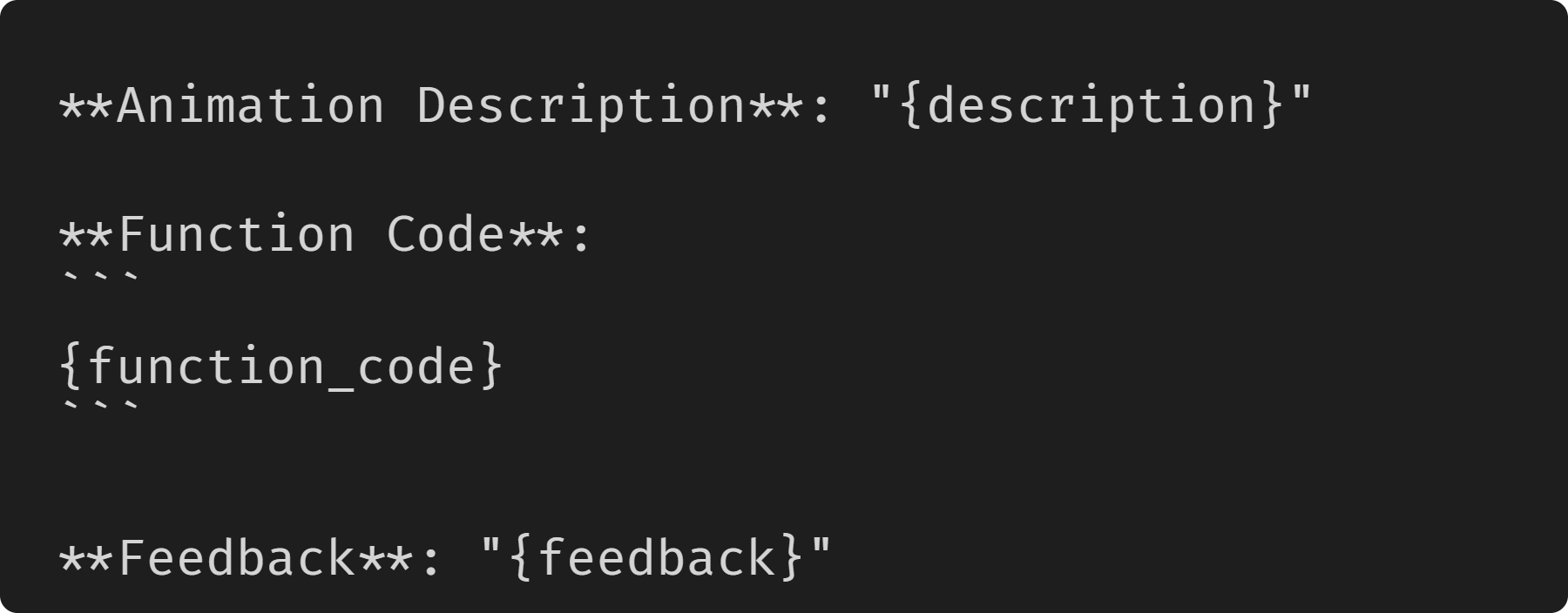}
    \caption{User message template for refining Gaussian transformations based on user feedback.}
    \label{fig:feedback-user}
\end{figure}

\subsection{Prompts Used in the Figures}

Below, we list the exact textual prompts used to generate animations presented in our experiments. These inputs directly guide animation generation for our method and the baselines.

\noindent \textbf{"Turn the vase into lava.", "Melting like lava."}
\begin{itemize}
    \item \textbf{Ours:} The animation can be broadly described as "Melting like lava". More detailed the centers slowly slide downwards like a thick liquid, but starting from the outer shell of the object towards also the core liquidating. Meanwhile, the colors change slowly to lava colors and opacities can stay unchanged. IMPORTANT: The Z coordinate for all center does not end up lower than the lowest Z of the original centers. This will make it seem as if the object is standing on the ground and the lava just lays on the ground at the end.
    \item \textbf{Baselines (Gaussians2Life~\cite{wimmer2025gaussianstolife} \& AutoVFX~\cite{hsu2024autovfx}):} The animation can be broadly described as melting the vase with flowers like lava. More detailed the centers slowly slide downwards like a thick liquid, but starting from the outer shell of the object towards also the core liquidating. Meanwhile, the colors change slowly to lava colors and opacities can stay unchanged. IMPORTANT: The Z coordinate for all center does not end up lower than the lowest Z of the original centers. This will make it seem as if the object is standing on the ground and the lava just lays on the ground at the end.
\end{itemize}

\noindent \textbf{"The bulldozer accelerates forward."}
\begin{itemize}
    \item \textbf{Ours \& Gaussians2Life~\cite{wimmer2025gaussianstolife}:} The bulldozer accelerates forward.
    \item \textbf{AutoVFX~\cite{hsu2024autovfx}:} The baseline failed to generate a valid function with the initial prompt. To work around this limitation, we attempted two alternative prompt variations: 
    \begin{itemize}
        \item The bulldozer accelerates forward. Do not put numpy arrays into animation requests.
        \item The bulldozer accelerates forward to (0, 1, 0).
    \end{itemize}
    However, neither of these resulted in a successful animation.
\end{itemize}

\noindent \textbf{"Shift through the colors of the rainbow"}
\begin{itemize}
    \item \textbf{All Methods:} The object stays stationary, but the colors shift through all colors of the rainbow. At the end, the object should transition back to its original colors to create a loop animation.
\end{itemize}

\noindent \textbf{"An explosion like a powder keg"}
\begin{itemize}
    \item textbf{All Methods:} An explosion like a powder keg.
\end{itemize}

\noindent \textbf{"The bear expands and contracts like a lung."}
\begin{itemize}
    \item \textbf{Ours:} A breathing animation, where the object expands and contracts like a lung.
    \item \textbf{Baselines (Gaussians2Life~\cite{wimmer2025gaussianstolife} \& AutoVFX~\cite{hsu2024autovfx}):} A breathing animation where the bear expands and contracts like a lung.
\end{itemize}

\section{Additional Qualitative Results}
\begin{figure*}[t]
    \centering
    \includegraphics[width=\linewidth]{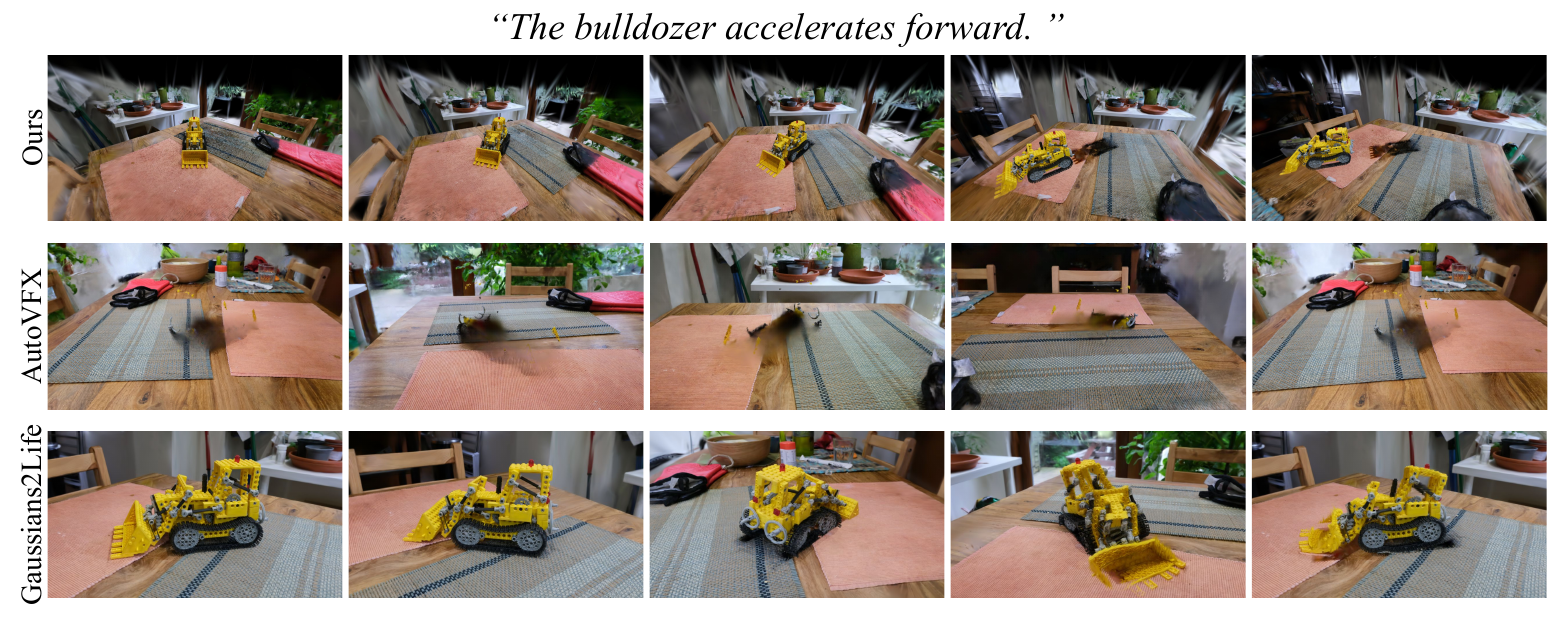}
    \caption{{\bf Qualitative comparison} with baselines on the lego bulldozer scene. Solely our method is able to arrive at an animation that fits the animation description. Furthermore it showcases impressive capabilities in object displacement tasks.} 
    \label{fig:additional-qualitative_comparison}
\end{figure*}

We provide additional visual comparisons against baseline methods in \autoref{fig:additional-qualitative_comparison}. Specifically, we demonstrate our method’s superior alignment to the animation prompts, particularly highlighting its capability in accurate object displacement, compared to AutoVFX~\cite{hsu2024autovfx} and Gaussians2Life~\cite{wimmer2025gaussianstolife}.

\end{document}